\documentclass[
 reprint,
 amsmath,
 amssymb,
 aps,
]{revtex4-2}
\usepackage{booktabs}
\usepackage{subcaption}
\usepackage{graphicx}
\usepackage{dcolumn}
\usepackage{bm}
\usepackage{braket}
\usepackage{hyperref}
\usepackage{diagbox}
\usepackage{overpic}
\usepackage{xcolor}
\usepackage{orcidlink}
\usepackage{cleveref}
\usepackage{diagbox}
\usepackage{float}

\begin{document}

\title{Distinct Modes of Quantum Information Transfer in Power-Law Long-Range Spin Networks}

\author{E.E. Marshall$^{1,\dagger}$\orcidlink{0009-0004-7866-7777}, C.C. Nelmes$^{1,2,*,\dagger}$\orcidlink{0009-0002-1686-6282}\thanks{Correspondence to: c.nelmes@york.ac.uk and irene.damico@york.ac.uk.}, T.J.G. Apollaro$^3$\orcidlink{0000-0002-9324-9336}, T.P. Spiller$^{1,2}$\orcidlink{0000-0003-1083-2604}, and I. D'Amico$^{1,2}$\orcidlink{0000-0002-4794-1348}}\thanks{Correspondence should be addressed to: c.nelmes@york.ac.uk and irene.damico@york.ac.uk.}

\affiliation{$^1$School of Physics, Engineering and Technology, University of York, York, YO10 5DD, United Kingdom}

\affiliation{$^2$York Centre for Quantum Technologies, University of York, York, YO10 5DD, United Kingdom}

\affiliation{$^3$Department of Physics, University of Malta, Msida MSD 2080, Malta}

\affiliation{$^\dagger$These authors contributed equally to this work}

\date{\today}

\begin{abstract}
We identify different regimes of quantum state transfer in long-range coupled spin-$\frac{1}{2}$ systems, where naturally occurring power-law interactions enable rapid, high-fidelity transfer with minimal engineering. Across a broad range of interaction profiles, from effectively nearest-neighbour coupling to Coulomb interactions, we show how long-range connectivity fundamentally reshapes the mechanisms underlying information propagation within such systems. For effectively short-range interactions, transfer follows familiar ballistic transfer dynamics: an initially localised excitation spreads across many eigenmodes concentrated within the approximately linear region of the spectrum, enabling robust wavepacket motion. In contrast, increasing long-distance interactions via lowering the power-law exponent $\alpha$  ($\alpha=1-2$) drives a striking transformation, where the initial state becomes confined to progressively fewer eigenmodes, ultimately reducing the dynamics to the coherent participation of only a few states corresponding to the highest eigenenergies. This spectral localization gives rise to emergent long-range oscillations between distant sites, revealing a distinct -- and faster -- transfer mechanism arising from the intrinsic structure of long-range quantum interactions rather than from full-system engineering pathways.
\end{abstract}

\maketitle


\section{\label{intro}Introduction}
Following the seminal proposal of quantum-advantaged information transfer through unmodulated (Heisenberg) spin chains by Sougato Bose \cite{Bose_2003}, and the introduction of perfect state transfer (PST) via engineered mirror-symmetric systems \cite{Christandl_2004,Nikolopoulos_2004}, significant effort has been devoted to improving both the speed and robustness of quantum communication via solid-state hardware. Along with a plethora of other quantum information processing tasks through the natural dynamics of spin chains \cite{Bose_2006,Kay_2010,Niko_2014,PhysRevA.84.032308,PhysRevA.95.042335,Alsulami2022}, quantum state transfer holds particular applicability to the scaling of quantum computing hardware. PST protocols enable exact state mirroring in fully engineered spin networks \cite{Christandl_2004,Albanese_2004,Karbach_2005,Kay_2006,Christandl_2017, Asoudeh2025}, while ballistic transfer schemes \cite{Apollaro_2012,Zwick_2012,Faria_2025} enable high-fidelity/quasiperfect state transfer (QPST) at times comparably faster than times required for total state mirroring in exclusively nearest-neighbour (NN) settings \cite{Yung_2006}. These ballistic approaches exploit an encoding of quantum information as a wavepacket in $k$-space \cite{Osborne_2004}. They typically require only boundary-coupling engineering and local magnetic field control to position the wavepacket within the linear region of the energy spectrum. Previous studies have shown that a linear energy spectrum enables the fastest perfect quantum state transfer \cite{Nikolopoulos_2004,Christandl_2004,Yung_2006,Kay_2010,Nelmes_2026}. In addition to ballistic wavepacket propagation, minimalist engineering approaches that tune none or at most a few couplings or on-site energies can give rise to few-mode coherent oscillations between the end sites \cite{PhysRevA.72.034303,Lorenzo_2013,PhysRevA.95.042335,Chetcuti_2020}, enabling QPST, although this generally comes at the cost of very long transfer times.

Despite the natural emergence of power-law long-range (LR) interactions in promising hardware platforms for fault-tolerant universal quantum computation \cite{Cirac_1995,Loss_1998,Deutsch2000}, their potential as quantum buses has received comparatively little attention over the past two decades, with only a limited number of studies to date \cite{Kay_2006,Gualdi_2008,Hermes_2020,Ferron_2022,Lewis_2023,Li_2025,Ahuja_2026,raupach2026robusttopologicalquantumstate}. While extensive research has explored state transfer protocols in nearest-neighbour systems for high-fidelity transfer \cite{Bose_2003,Christandl_2004,Zwick_2011,Apollaro_2012,Zwick_2012,PhysRevA.95.042335,Alsulami2022,Lorenzo_2013,Giorgi_2013,Niko_2014,Riegelmeyer2021,Bezaz_2025,Nelmes_2026, Michel2026}, and more recently in next-nearest-neighbour (NNN) extensions \cite{Christandl_2017,Faria_2025}, long-range interacting models have historically received comparatively little attention. The NN and NNN approaches span fully engineered networks and minimally controlled schemes based on boundary-parameter tuning \cite{Zwick_2011,Apollaro_2012,Zwick_2012,Banchi_2011,Lorenzo_2013,Faria_2025}, both of which can achieve PST and/or QPST in XX spin chains. However, these techniques have not yet been systematically extended to genuinely long-range interacting systems, particularly in experimentally-accessible system sizes of $N \sim 10^2$ sites \cite{Cheng_2023,Daka_2026}.

In the present work, we focus on the physically relevant regime of experimentally-realistic interactions and system sizes, aiming to identify transfer protocols that remain both scalable and operationally viable within feasible coherence times. We focus on a range of values of the decay exponent $\alpha$ defining the power-law interaction, namely $\alpha\in [1,10]$. These values naturally arise, or can be effectively approximated, across a variety of quantum platforms \cite{Defenu_2023,Jurcevic2014,PhysRevApplied.17.034060}. 
\section{\label{model}Spin Model}
We introduce the spin-$\frac{1}{2}$ so-called XX (also XY isotropic) model
\begin{equation} \label{Hami}
    \hat{H}_{XX}=
\frac{1}{2}\sum_{i < j}
J_{i,j}
\left(\hat{\sigma}_i^x \hat{\sigma}_j^x + \hat{\sigma}_i^y \hat{\sigma}_j^y \right)
+
\frac{1}{2}\sum_i B_i \hat{\sigma}_i^z
\end{equation}
where
\begin{equation}
J_{i,j}=\frac{J_0}{r_{i,j}^{\alpha}}
\label{Coupling}
\end{equation}
are the site-dependent couplings defined by the dimensionless relative positions $r_{i,j}=|i-j|$ between sites, $J_0$ is the maximal bulk coupling energy, and $B_i$ are the site-specific magnetic fields/on-site energies. Clearly, as we increase $\alpha$ within Equation (\ref{Coupling}), it can be seen that Equation (\ref{Hami}) tends towards nearest-neighbour (NN) interactions, and as we decrease it, we tend towards full network connectivity. The power-law exponents that commonly arise in quantum systems are $\alpha = 1$ for Coulomb-mediated trapped-ion spin models \cite{Welzel_2011,Johanning2016}, $\alpha = 3$ for magnetic or electric dipole–dipole interactions \cite{Barredo,PhysRevApplied.17.034060,Defenu_2023}, and $\alpha = 6$ for van der Waals interactions \cite{Adams_2020,Zhao2023}. In this study, we consider values of $\alpha$ from 1 to 10 to elucidate how different effective interaction ranges influence the state transfer dynamics. As an example, trapped-ion platforms allow the power-law exponent to be tuned experimentally \cite{Richerme2014,Defenu_2023}.
\section{Quantum State Transfer}\label{QST}

\begin{figure}
    \centering
    \includegraphics[width=1.0\linewidth]{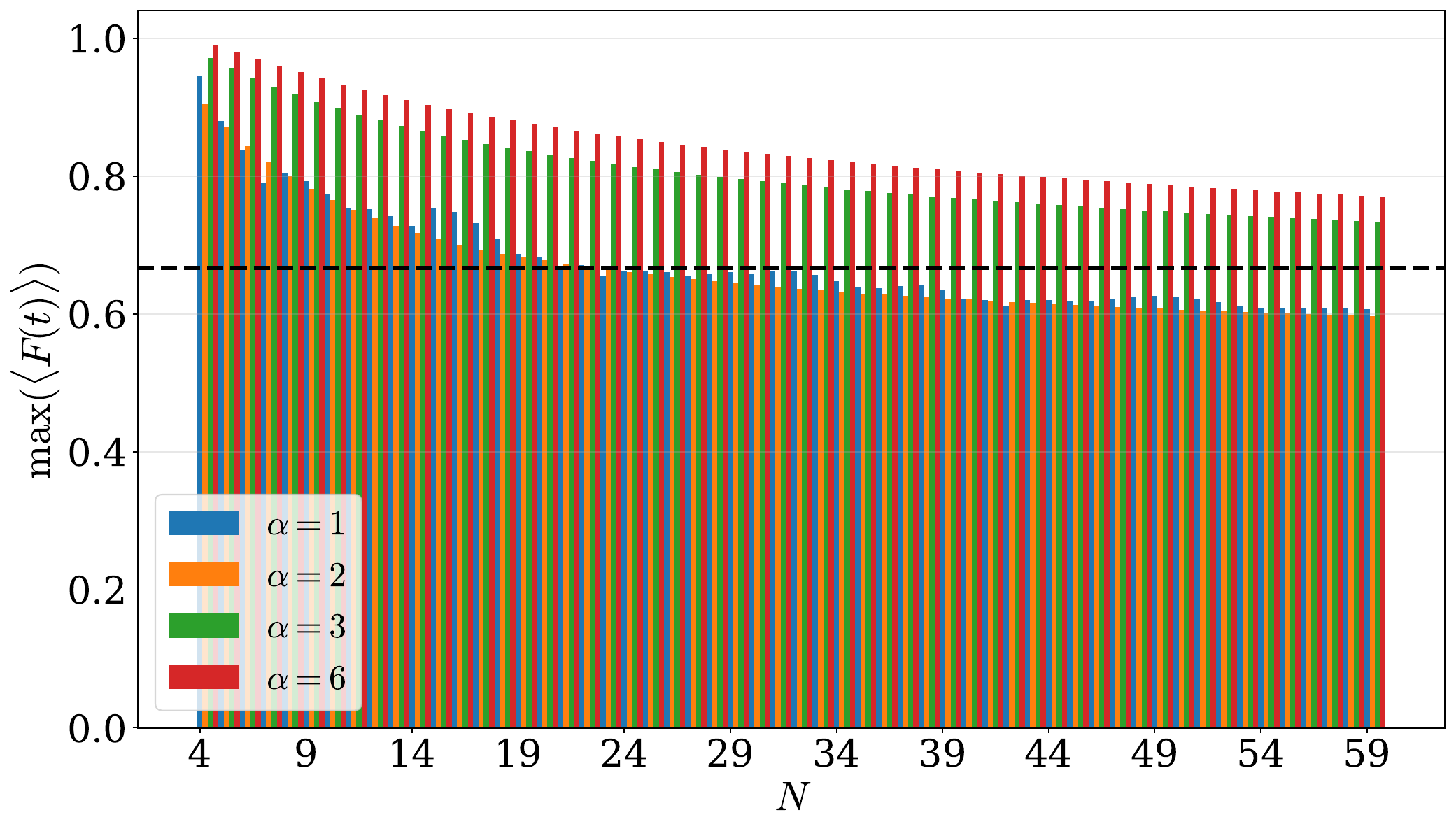}
    \caption{Maximal average fidelity as a function of total number of sites $N \in [4,60]$ for $\alpha$-chains, evaluated over a total time window $t \cdot J_0 = N\pi/2$, corresponding to twice the speed limit for perfect state transfer (PST) in nearest-neighbour systems \cite{Yung_2006}. Horizontal (dashed) line corresponds to the limit in which there is quantum advantage over classical means of communication \cite{Horodecki}. In the large-$\alpha$ limit, the XX (effective) nearest-neighbour spin chain is known to lose its quantum advantage for communication at $N \approx 240$ \cite{Bayat_2011}.}
    \label{Fig_1}
\end{figure}

The quantum state transfer (QST) scheme considered here was originally introduced in Ref.~\cite{Bose_2003} and consists of encoding an arbitrary single-qubit state $
\ket{\psi_s}
=
\cos\frac{\vartheta}{2}\ket{0}
+
\sin\frac{\vartheta}{2}e^{i\phi}\ket{1}$
at a designated sender site. The state is subsequently transferred through the natural unitary dynamics of the spin chain and recovered at a different location, called the receiver site, after a time \(\tau\). Throughout this work, the sender and receiver are taken to be the opposite ends of the chain, namely sites \(1\) and \(N\), respectively. Let $|\Psi(t)\rangle=e^{-i\hat{H}_{XX}t}|\Psi(0)\rangle$ denote the state of the entire spin chain at time $t$. The reduced density matrix of the receiver spin is
$\hat{\rho}_r(t)
=
\operatorname{Tr}_{1,\ldots,N-1}
\left[
|\Psi(t)\rangle\langle\Psi(t)|
\right].$ The performance of the QST protocol is quantified via the fidelity between the initial sender state and the state reconstructed at the receiver site, averaged over all pure input states
\begin{align}
    \label{eq:averageF}
    \braket{F(t)}=\frac{1}{4\pi}\int d\Omega \bra{\psi_s}\hat{\rho}_r(t)\ket{\psi_s}~.
\end{align}
Initialising all but the sender spin in the fully polarised state, i.e., $\ket{\Psi(0)}=\ket{\psi_s}_1\otimes_{i=2}^{N}\ket{0}_i$, we can exploit the model's $U(1)$-symmetry to restrict the dynamics to the single-particle preserving subspace, yielding from Equation (\ref{eq:averageF}) \cite{Bose_2003}
\begin{align}
    \label{eq:averageFB}
    \braket{F(t)}=\frac{1}{2}+\frac{\left|f_1^N(t)\right|}{3}+\frac{\left|f_1^N(t)\right|^2}{6}~,
\end{align}
where $f_1^N(t)=\bra{N}e^{-i \hat{H}_{XX} t}\ket{1}$ is the single-particle transition amplitude from site 1 to site $N$, adopting the notation $ 
    \ket{i}
    =
    \ket{0_1}\otimes\cdots\otimes
    \ket{1_i}
    \otimes\cdots\otimes
    \ket{0_N}$, with $  i \in \{1,\ldots,N\}$. Hence, the average fidelity $\braket{F(t)}$ is a monotonically increasing function of the absolute value of the transition amplitude, which is given by
\begin{align}
    f_1^N(t)=\sum_{k=1}^N v_{1k}v_{Nk}^*e^{-i \omega_k t}~,
    \label{wavetran}
\end{align}
where $\{\omega_k, \ket{v_k}\}$, with $\ket{v_k}=\ket{v_{1k},v_{2k},\cdots,v_{Nk}}$, are the eigenvalues and eigenvectors of $\hat{H}_1$, the Hamiltonian in the single-particle sector.  Clearly, when Equation~(\ref{eq:averageF}) equals unity at some time $\tau$ where the fidelity is maximal, this signifies PST. Any spin chain configuration which yields an average fidelity value at $\tau$ less than unity but still $\gtrsim 0.99$ is classified as QPST. Figure~\ref{Fig_1} shows how the state-transfer fidelity of completely unmodulated long-range spin chains with power-law interactions, described by Equation~(\ref{Hami}), deteriorates with increasing system size $N$ for different values of the interaction exponent $\alpha$. This demonstrates that some sort of (preferably minimal) intervention or modulation is required, to deliver QPST from such chains as $N$ is increased. This is the motivation for the work presented here.

\label{NNN}
\section{Optimisation Approach}\label{methods}
We employ evolutionary computation methods for this investigation. In particular, genetic algorithms (GAs) \cite{H1992, E2015, L2019, Feng2019,Alhijawi2024} operate by initialising a population of candidate solutions (“individuals”), each encoded by a set of parameters that determine their performance with respect to a given objective. This population is then evaluated through a fitness (or cost) function that quantifies how well each individual performs within a specific environment. Individuals with higher fitness are preferentially selected to generate offspring through genetic operators such as crossover and mutation, thereby exchanging and modifying information across candidate solutions. The resulting offspring constitute a new generation and are reintroduced into the same environment for further evaluation.
This iterative process is repeated over successive generations, progressively refining the population toward higher fitness. Termination is typically determined either by convergence when no significant improvement in the best fitness is observed or upon reaching a predefined maximum number of generations.

\begin{figure}
    \centering
    \includegraphics[width=\linewidth]{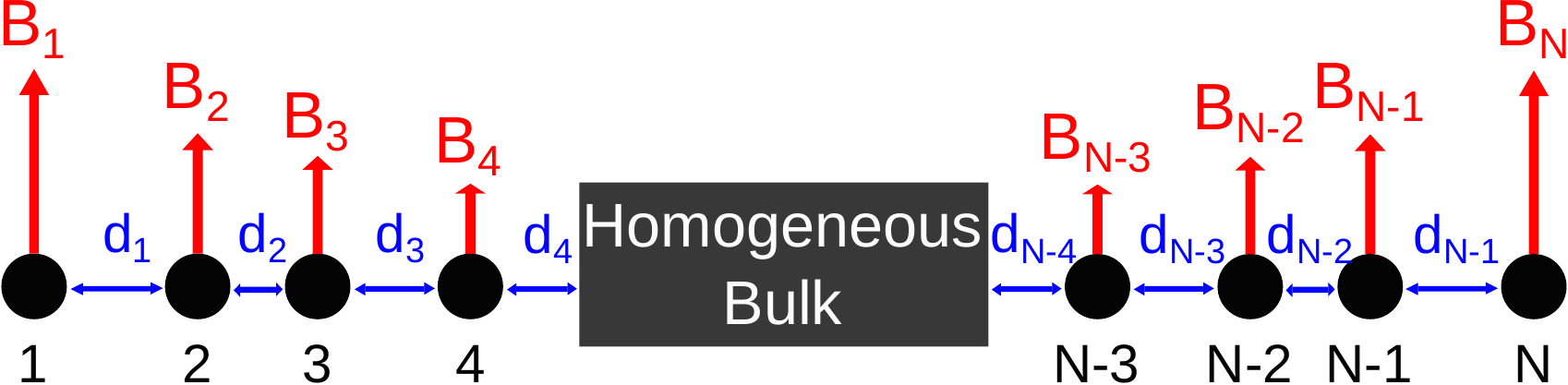}
    \caption{A diagram showing an example chain of length N with 4 optimised site positions, with $d_i$ the distance between two adjacent sites   and on-site energies, $B_i$, on each end of the chain, preserving mirror symmetry.}
    \label{fig:Chain-diagram}
\end{figure}

GAs have been successfully employed to identify novel Hamiltonian configurations for the optimisation of information processing tasks \cite{Bezaz_2025,LM2021,PS2025,Faria_2025}. In fact, recent studies suggest that genetic algorithms may offer advantages in convergence time compared with other mainstream optimisation strategies \cite{PS2025,santana_2026}, particularly in noiseless settings. The specific algorithm implemented in this study was adapted from that which was used in Refs.~\cite{LM2021,Bezaz_2025,Faria_2025}, in order to assess minimalist engineering protocols within power-law LR models. The fitness function for the specific algorithm employed within this study is
\begin{eqnarray}
x(f_1^N(\tau),\tau) = A \ \text{e}^{(a(|f_1^N(\tau)|^2-1))},
\label{ff}
\end{eqnarray}
where $A=100$ and $\tau$ is the time at which maximum excitation transfer fidelity is attained. The time $\tau$ (in units of $J_{0}^{-1}$) is within the time interval of $t\cdot J_{0}\in[0,\frac{N\pi}{4}]$, where $\frac{N\pi}{4J_0}$ is the speed limit for NN chains \cite{Yung_2006}.
The algorithm commences via an initialisation of a population of randomly generated individuals of long-range fixed-alpha chains, as described by Equation~\eqref{Hami}, where the bulk of the chain spacing is homogeneously spaced, and provides the referential spacing set to unity. Depending on the number of sites that are being optimised, the algorithm initialises a random configuration beginning from the outer sites, maintaining mirror symmetry. Therefore, if there is only one site being optimised from the beginning of the chain, then one position and local magnetic field is randomly generated and then subjected to the fitness evaluation via the fitness function Equation~(\ref{ff}). Note that in order to vary the couplings of one (or more) end sites, whilst maintaining the chosen LR power-law exponent, the dimensionless position of that site is varied, to change all the relevant couplings via Equation~(\ref{Coupling}). As mirror symmetry is enforced throughout the mutation process, this assesses also the identical position and magnetic field profile on the opposite side of the chain. An example chain with 4 optimised energies and spacings is depicted in Figure \ref{fig:Chain-diagram}, with the optimisations mirrored to the opposite end of the homogeneous bulk. For this study, the site positions were constrained to an inter-site distance $d_i$ within the range
$d_i \in [0.5, 5.0]$, while the local magnetic fields were restricted to
$B_i \in [0, 10 J_0]$. These bounds were selected to remain within
experimentally plausible parameter regimes, where site-resolved on-site energy modulations can be engineered using magnetic-field gradients \cite{Wolk_2017}, external harmonic potentials \cite{Weitenberg2011}, or specific tuning of optical superlattice potentials \cite{Atala_2013}. We also enforce the bounds of the positions' variability to ensure that no edge couplings exceed the characteristic energy (bulk coupling) of the system by more than an order of magnitude through adjacent sites becoming arbitrarily close to one another. 
\begin{table}
\scriptsize
\centering
\caption{Genetic algorithm parameters used for the optimisation of spin-chain configurations for high-fidelity state transfer.}
\label{tab:GA_params}

\begin{tabular}{|l|c|}
\hline
\hline
Parameter & Value \\
\hline
Selection method & Roulette-wheel selection \\
Crossover method & Uniform crossover \\
Initial mutation rate & 20\% \\
Final mutation rate & 1\% \\
Generations & 100 \\
Genomes per generation & 4096 \\
Time divisions per unit time & 100 \\
Stopping criterion & Maximum generations reached \\
Reported value & Best genome found within individual runs \\
\hline
\hline
\end{tabular}

\end{table}

Once the fitnesses have been evaluated and recorded, a crossover function exchanges 50\% of the genetic encoding of each of the parents to create the offspring. This process allows for the traits which might allow for the highest-fidelity transfer to be passed on to the next generation of prospective solutions.
The initial genomes are defined as uniform power-law LR chains with each genome in the population pool then mutated five times to generate a random initial population. During each mutation, a position within the genome is selected at random (within the initial bounds), and a random integer less than or equal to the maximum mutation rate is generated. This value is then either added to or subtracted from the selected position, with the maximum mutation rate decreasing with generation number, as specified in Table~\ref{tab:GA_params}. Consequently, if only a single site is being optimised from the beginning of the chain, only that position, and therefore its corresponding on-site energy, is randomly mutated before the resulting genome is evaluated using the fitness function given in Eq.~\ref{ff}. Iteration over these individuals across a sufficiently large number of generations and populations allows the optimisation to converge towards high-performing solutions, with convergence assessed once the stopping criterion of a maximum number of generations is reached. The fine tuning of the mutation rate, particularly one that decreases over generations, is advantageous for finding more precise regions of local optima. 
\section{Results}
Here, we present the results from the optimisation approach described by the previous section. For fair comparison across ranges of $N$ and $\alpha$, we set $J_0$ =1 as the characteristic energy scale for the dynamical analysis.
\subsection{Distinct Modes of transfer}

\begin{figure}
    \centering
    \includegraphics[width=0.9\linewidth]{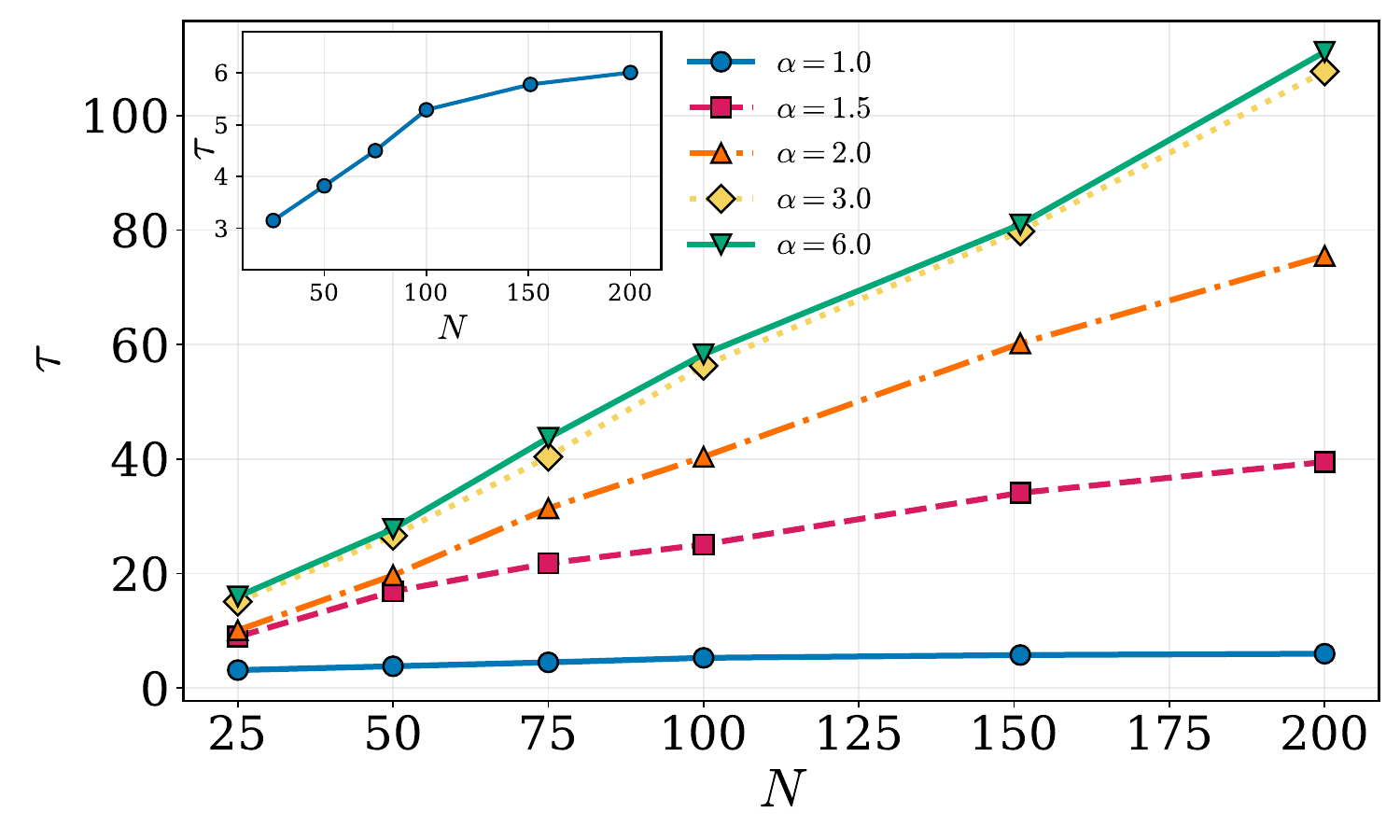}
    \caption{Time at which the fidelity is maximal $\tau$ plotted as a function of the total number of sites $N$, across increasing strengths of the power-law exponent $\alpha.$ Inset presents a zoomed-in window of the $\alpha=1$ chains, highlighting the distinguishability of the individual data points.}
    \label{fig_time}
\end{figure}

\begin{figure}[h!]
    \centering
    
    \includegraphics[width=0.9\linewidth]{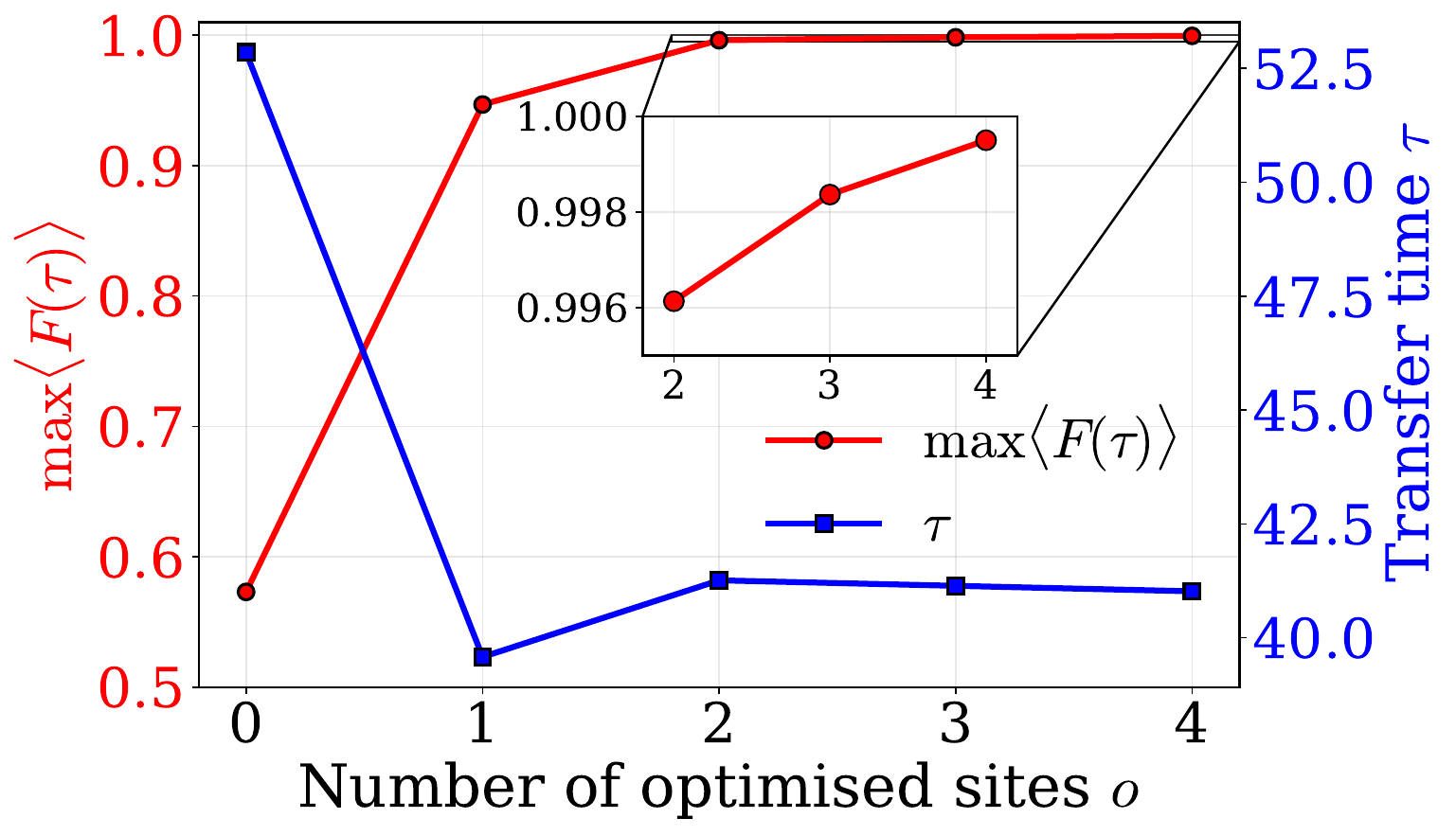}
    \includegraphics[width=0.9\linewidth]{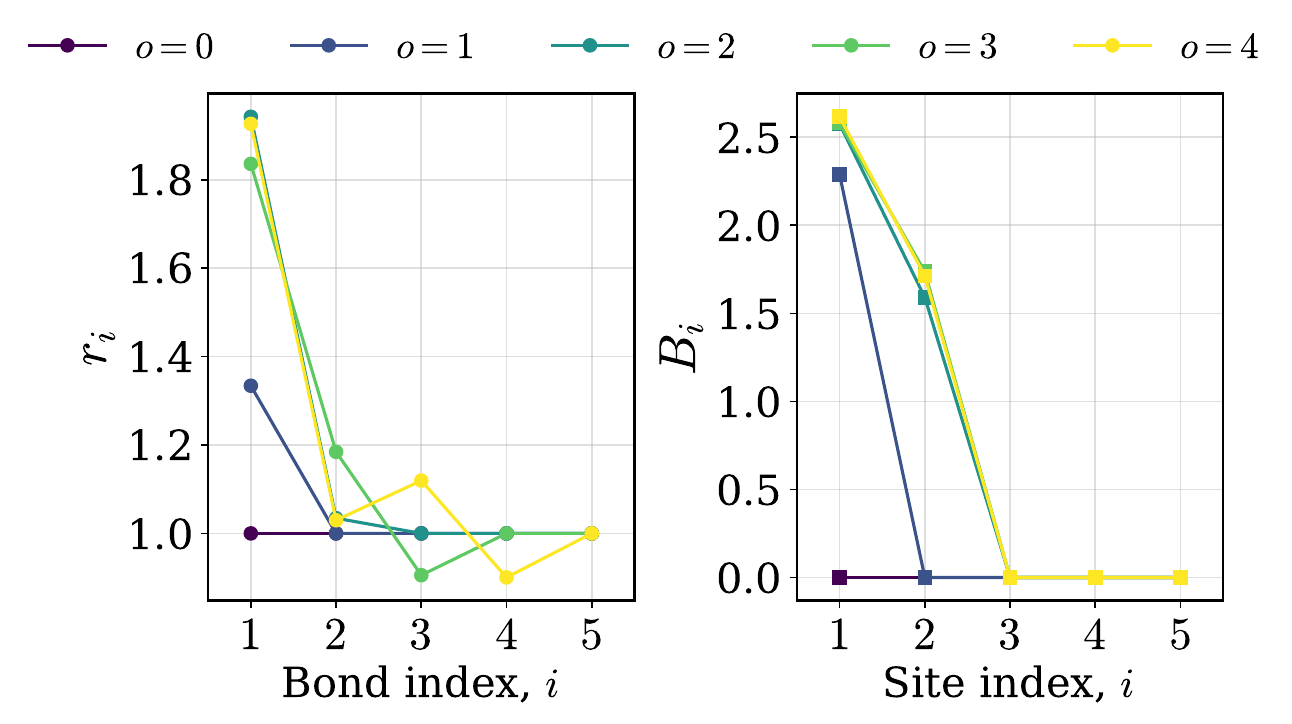}

    \caption{Top panel shows, for a fixed $N = 100$ site long-range ($\alpha = 2$) chain, how increasing the number of optimised boundary sites $o$, from 0 to 4 leads to near-perfect (~$99.99\%$) average fidelity and fast transfer time, eventually approaching a saturation regime. Bottom panel illustrates the corresponding optimised spatial profiles of the spacing between sites and local magnetic fields across the lattice: the bulk remains homogeneous with uniform bond strength and zero field modulation, while non-trivial structure develops only near the boundaries.}
    \label{fig_2}
\end{figure}

Table~\ref{tab:alpha_scan} summarises the QPST performance achieved using the optimised boundary-coupling schemes for system sizes $N = 10^{1} - 10^{2}$. For the great majority of systems, a maximum average fidelity of 0.999 is reached. Figure~\ref{fig_time} plots $\tau$ as a function of $N$ across the varied $\alpha-$systems, demonstrating the scaling of the time at which the average fidelity achieves its maximal value for the optimised solutions. This is sublinear for $\alpha\le 1.5$.
Figure~\ref{fig_2} shows how increasing the number of optimised sites for a fixed chain length and interaction exponent ($\alpha=2.0$) leads to saturation of both the maximum fidelity and the corresponding time $\tau$ at which it is attained. The bottom panel illustrates how modest modifications to the site positions and on-site energies are sufficient to achieve the highest fidelity for a given number of optimised boundary sites $o$. Correspondingly, Figure~\ref{fig_3} (See Appendix) presents how the increase in number of optimised sites yields the appearance of more participatory eigenmodes concentrated within a linear region near the top of the energy-ordered spectrum.

\begin{figure}[t]
    \centering

    \begin{minipage}[t]{0.49\linewidth}
        \centering
        \includegraphics[width=\linewidth]{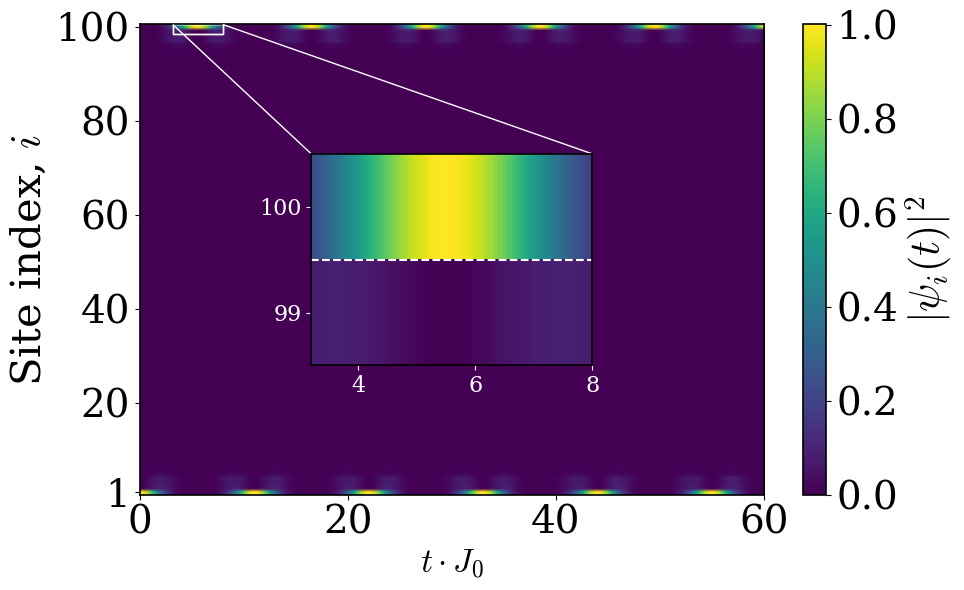}

        \vspace{-0.27cm}

        \includegraphics[width=\linewidth]{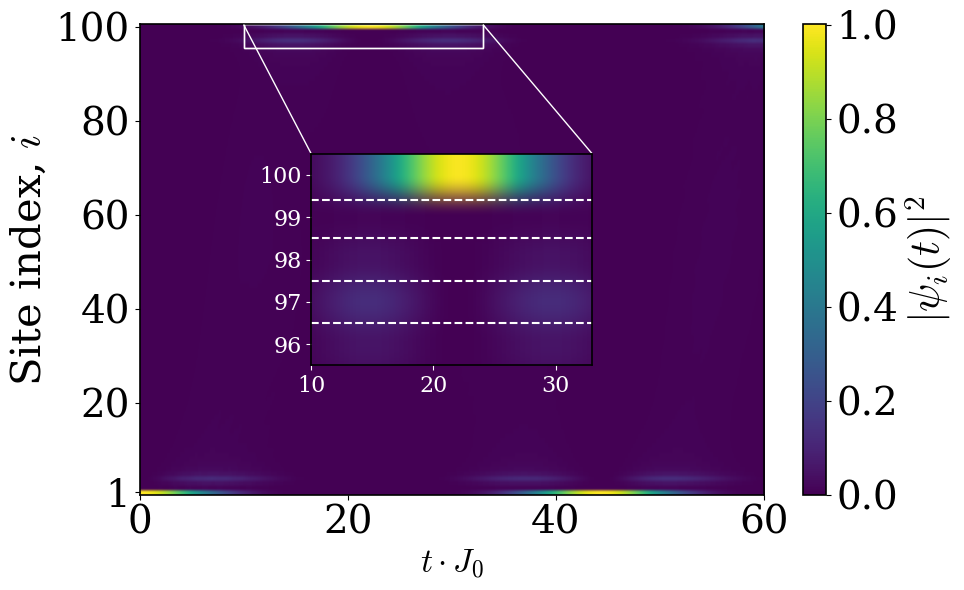}
        
        \vspace{-0.29cm}
        
        \includegraphics[width=\linewidth]{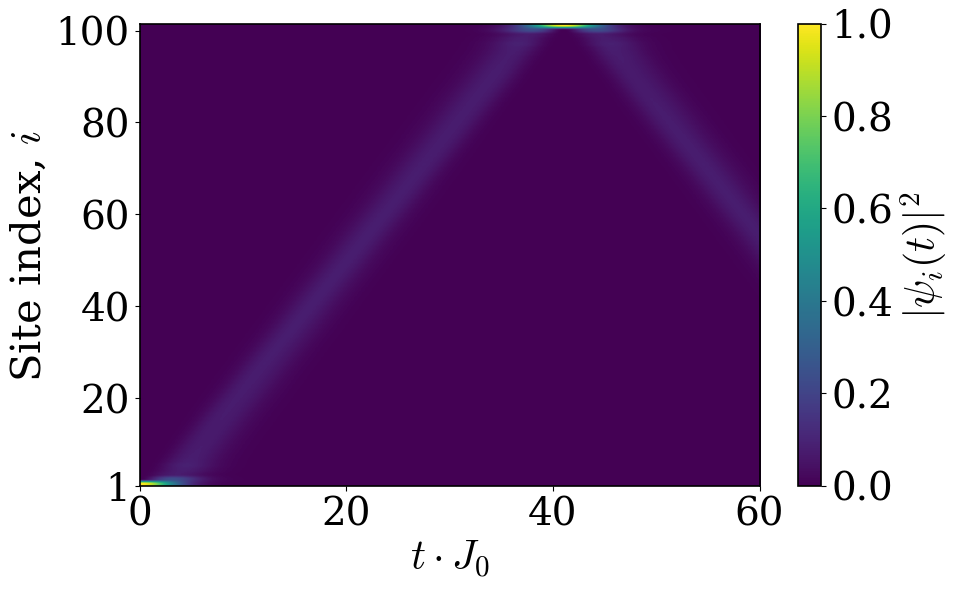}

        \vspace{-0.32cm} 

        \includegraphics[width=1.0\linewidth]{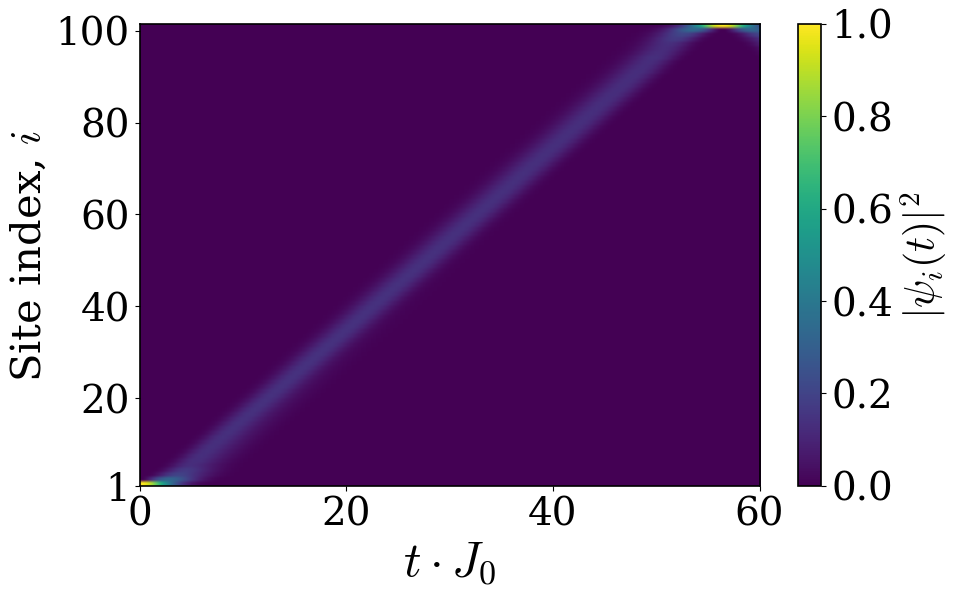}
    \end{minipage}
    \hfill
    \begin{minipage}[t]{0.49\linewidth}
        \centering
        \includegraphics[width=\linewidth]{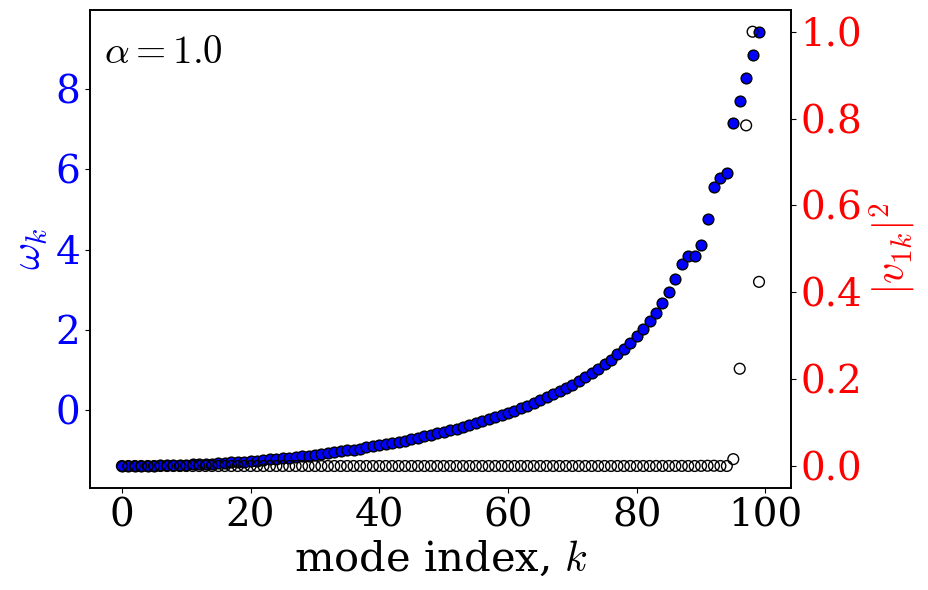}

        \vspace{-0.27cm}

           \includegraphics[width=\linewidth]{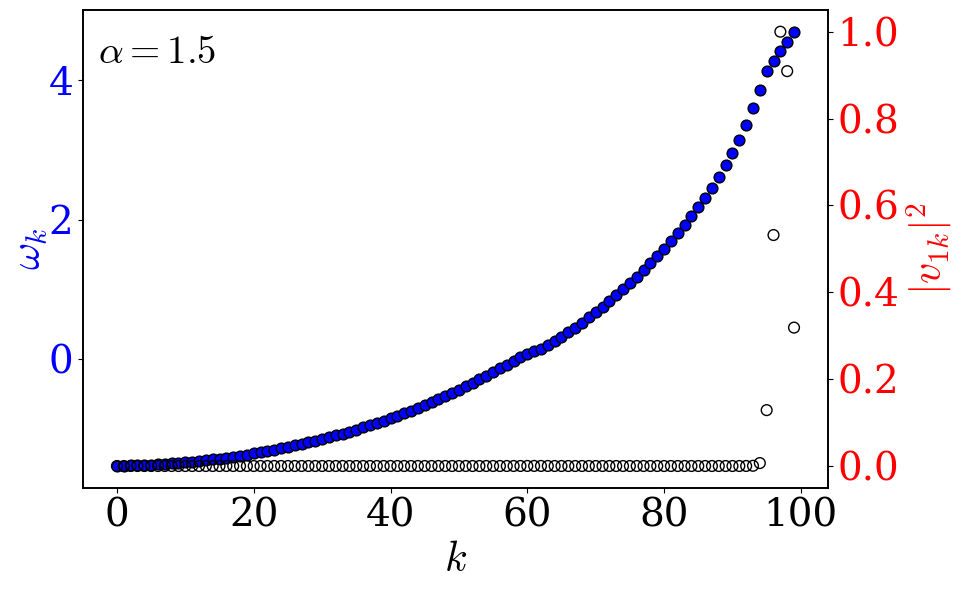}

        \vspace{-0.25cm}

        \includegraphics[width=\linewidth]{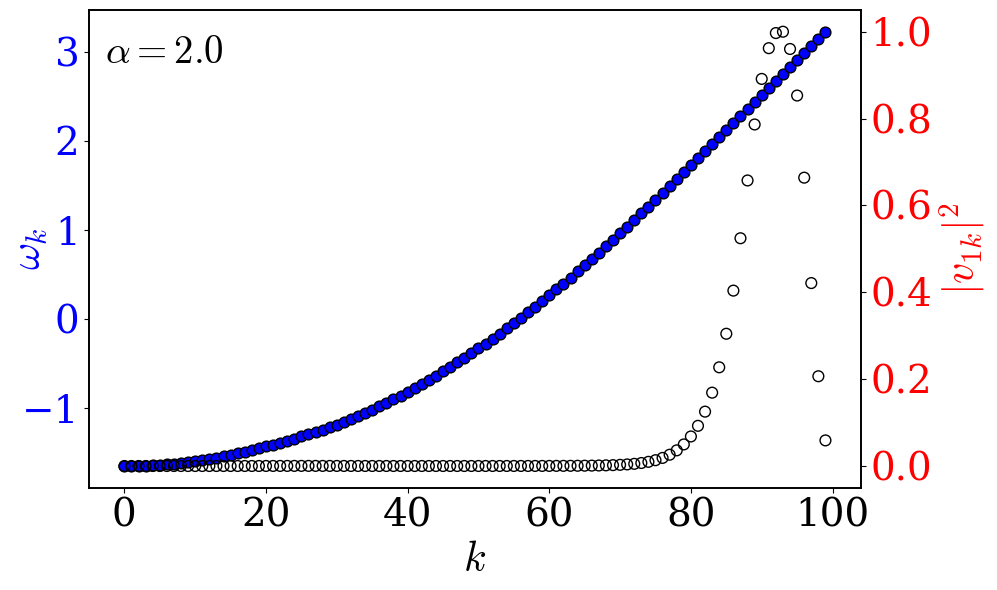}

        \vspace{-0.26cm}

        \includegraphics[width=\linewidth]{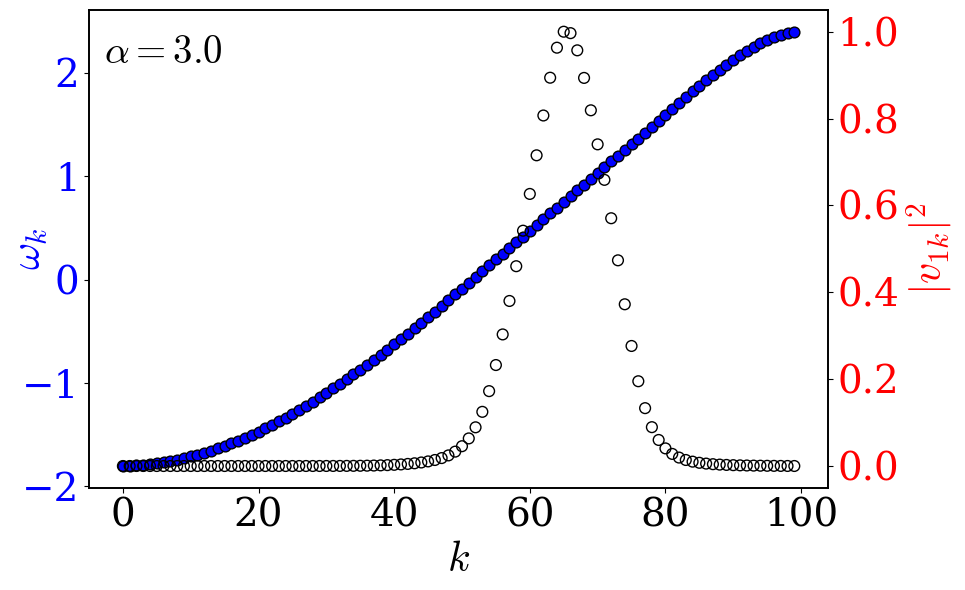}
    \end{minipage}

    \caption{Distinct transfer dynamics associated with optimized chains for the selected sequences of $\alpha$ values (left column), alongside the localization of the initial-state distribution in $k$-space over the ordered spectrum (right column) of the corresponding varied-$\alpha$, $N=100$-site chain.}
    \label{fig_4}
\end{figure}

Figure~\ref{fig_4} shows the dynamics, spectra, and initial-state distributions in $k$-space for different values of $\alpha$. The left column illustrates the spatial spread and localisation of the excitation as a function of time, while the right column shows the corresponding initial-state distribution and energy spectrum of the optimised solutions. A table detailing each configuration shown in Figure \ref{fig_4} is provided in Table~\ref{solutions} (See Appendix). For $\alpha=3.0$, the spectral concentration of the initial-state distribution and dynamics are highly consistent with both $\alpha=6.0$ (hence its omission) and previous ballistic transfer schemes \cite{Apollaro_2012}. Namely, this distribution localises, with many participatory-modes, in a mostly-linear segment of the ordered spectrum. This close agreement suggests that the transfer mechanism is somewhat ballistic in nature. To examine this, we consider the symmetry properties of the Hamiltonian's energy eigenvectors. Ballistic wavepacket transfer schemes exploit the mirror symmetry inherent to Hamiltonians that support high-fidelity quantum state transfer, for which the first and last components of the $k^{\text{th}}$ eigenvector satisfy $v_{1k}=(-1)^k v_{Nk}$~\cite{Cantoni1976}. Consequently, \cref{wavetran} can be rewritten as
\begin{align}
\label{eq:trans2}
f_1^N(t)=\sum_{k=1}^N v_{1k}^2 e^{-i \left(\omega_k t-k\pi\right)}~.
\end{align}
This expression admits a natural interpretation as a wavepacket whose spectral components are weighted by $v_{1k}^2$ and evolve according to the dispersion relation $\omega_k$. Following Refs.~\cite{Banchi_2011,Apollaro_2012,Faria_2025}, the ballistic wavepacket approach to achieve high-fidelity quantum state transfer is to engineer the spectral weight ($v_{1k}^2$) so that it is concentrated within a region where the ordered spectrum ($\omega_k$) is approximately linear. In this regime, all significant spectral components propagate with nearly identical phase velocities $\omega_k/k$, suppressing dispersion and enabling ballistic excitation transfer.

For $1\le\alpha\le 1.5$, in order to converge to the most optimal solutions, the total optimisation time window was progressively shortened: at difference with $\alpha\ge 1.6$, higher-fidelity solutions were found at significantly shorter transfer times, compared to other $\alpha$ solutions, within the prescribed bounds on the positions and magnetic fields. The resultant dynamics of these much faster solutions exhibit a markedly different regime of state transfer. This behaviour can be characterised as few-mode coherent oscillations between the end sites. This has some similarities to part of the nearest-neighbour system cases, where an effective Hamiltonian can emerge with only a few eigenstates entering the dynamics of Equation~(\ref{wavetran}) \cite{PhysRevA.72.034303,Lorenzo_2013,PhysRevA.95.042335}. In these cases, however, these few-states dynamics did not lead to fast transfer. Figure \ref{fig_4} shows that the $N=100$, $\alpha=1.0$ transfer is effectively mediated by just the highest four equally spaced energy eigenstates, with all others playing a negligible role. See Appendix~\ref{appendixc} for further discussion of this transfer mechanism underlying the $\alpha=1.0$ solutions. 

\begin{figure}[h!]
    \centering
    \includegraphics[width=\linewidth]{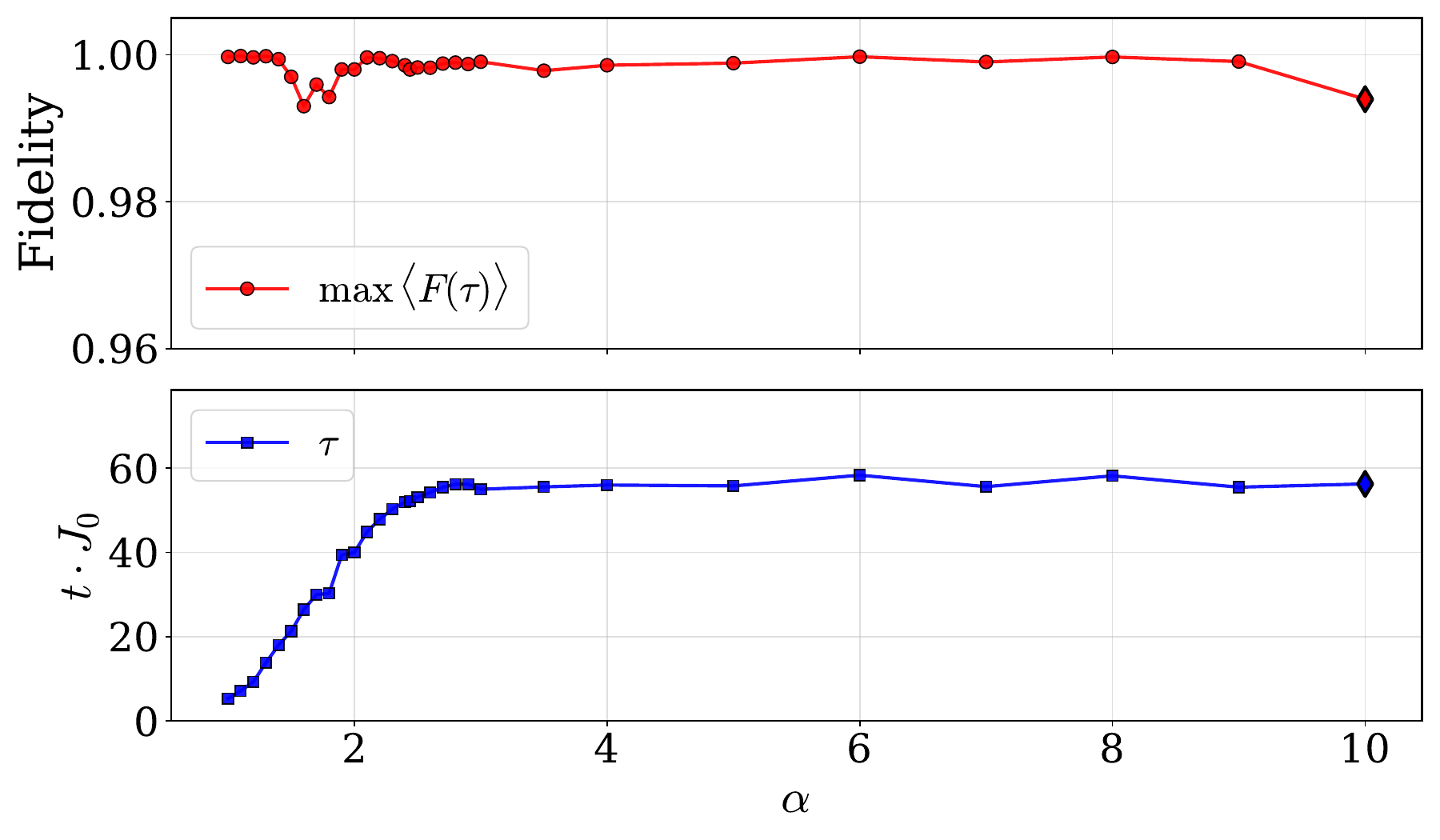}
    \caption{A plot of the maximal average fidelities and time taken for $N=100$ optimised chains for various $\alpha-$values between 1 and 10. Diamonds at the (right) end of the plots represent an overlap with the exact values of average fidelity, and time at which the fidelity is attained, for the exclusively nearest-neighbour case \cite{Apollaro_2012}.}
    \label{fig:Fidelity_plot}
\end{figure}

For $\alpha=1.5$, then on to $\alpha=2.0$, there is a growth in the number of energy eigenstates contributing to the transfer dynamics, with a wavepacket in $k-$space emerging, growing in width with increasing $\alpha$. Figures~\ref{fig_time} and \ref{fig:Fidelity_plot}, demonstrate that, for fixed $N$, the optimal transfer time $\tau$ is shortest for the lowest value of $\alpha=1.0$. The time grows with increasing $\alpha$ , as more energy eigenstates contribute and the wavepacket in $k-$space emerges and widens, up to approximately $\alpha=3$, when these growths effectively saturate. Notably, as shown in the upper panel of Figure~\ref{fig:Fidelity_plot}, the maximum fidelity remains always above $0.993$, with a small region for $1.0\leq\alpha\leq2.0$ in which the maximum fidelity decreases slightly. This is accompanied by a discontinuous increase in the time $\tau$ at which the maximum fidelity is attained. We suggest that this region is indicative of the transitional regime between the few-mode coherent oscillation mechanism and ballistic wave-packet propagation. 

\begin{figure}[t]
    \centering

    \begin{minipage}[t]{0.49\linewidth}
        \centering
        \includegraphics[width=\linewidth]{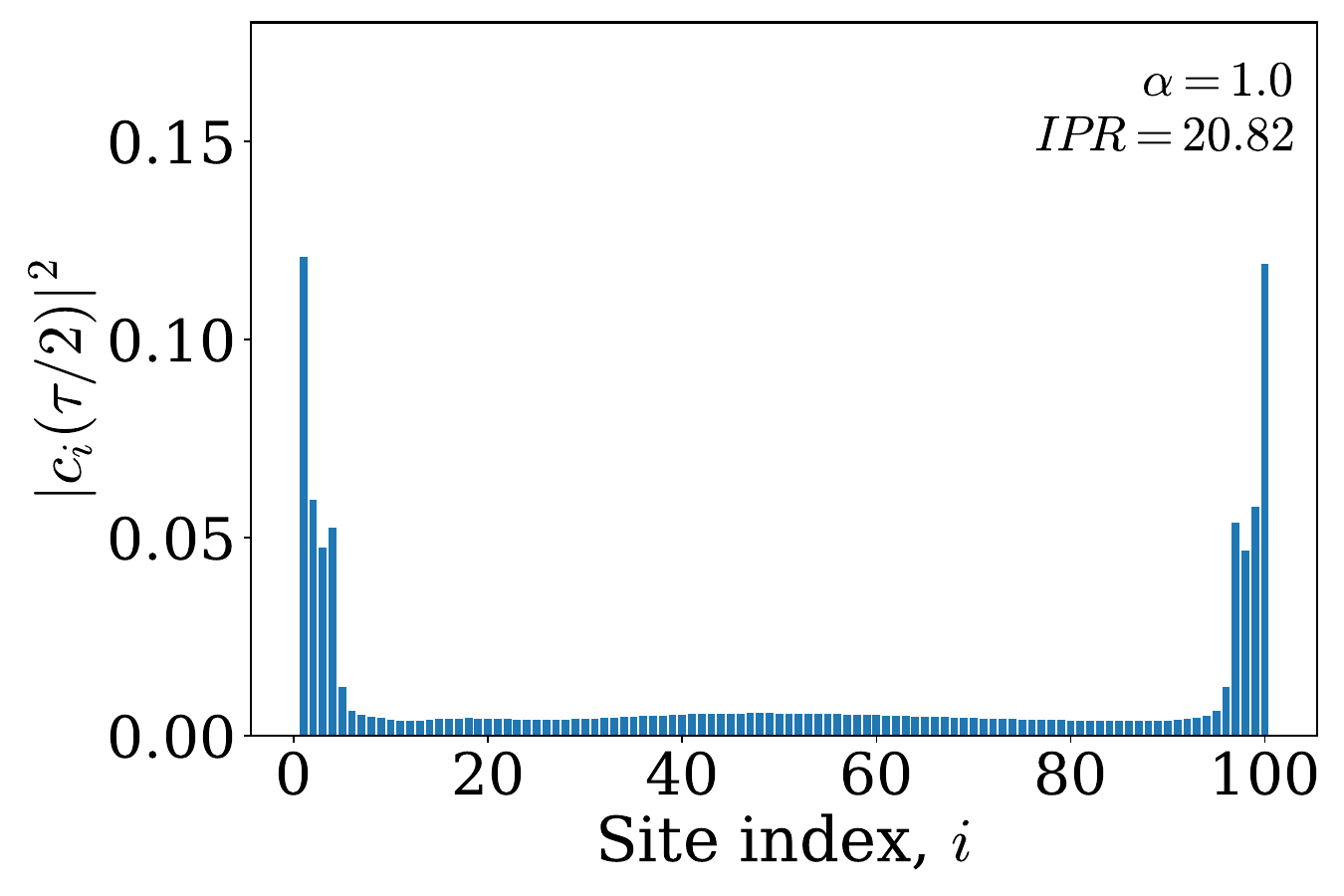}

        \vspace{-0.3cm}

        \includegraphics[width=\linewidth]{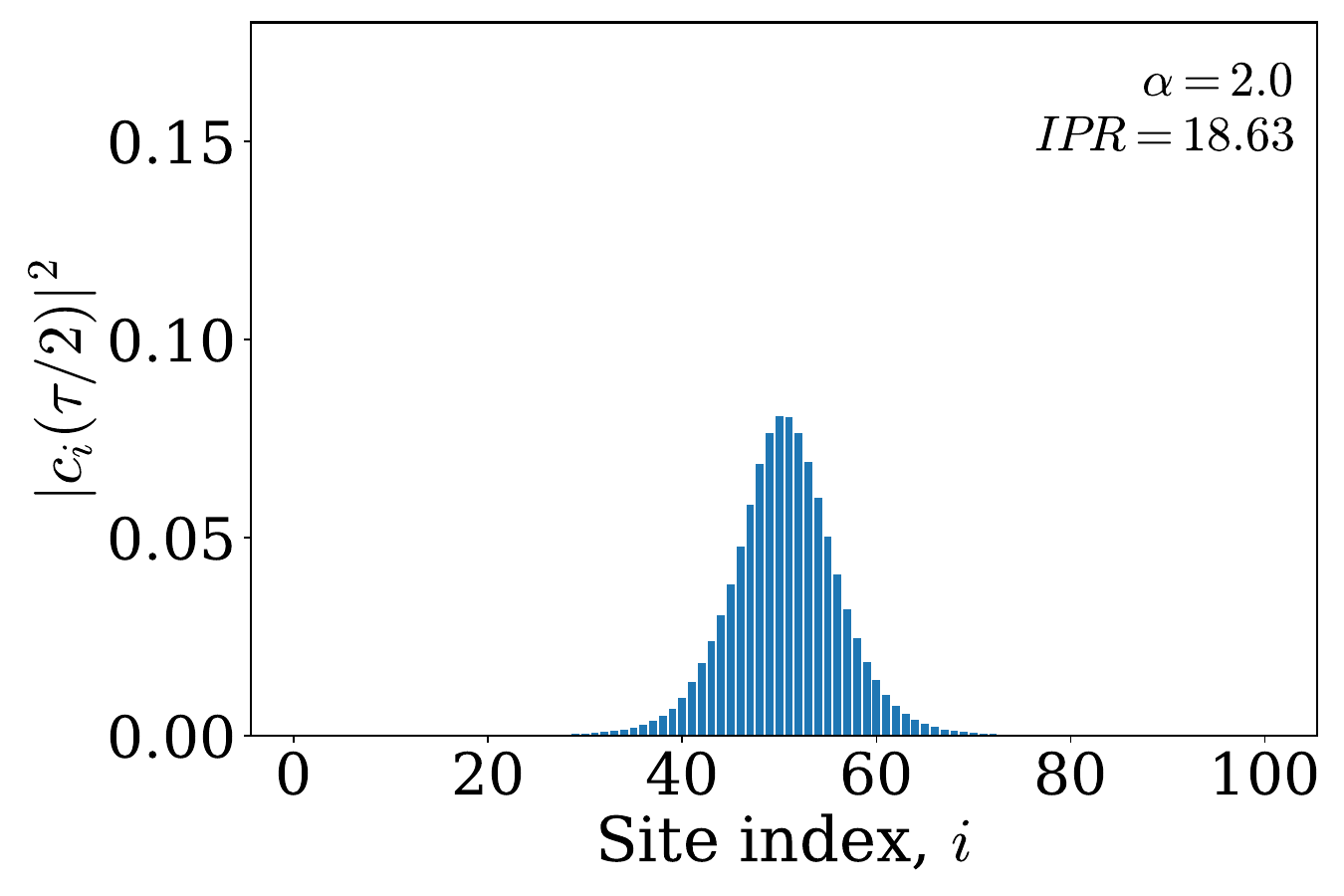}

    \end{minipage}
    \hfill
    \begin{minipage}[t]{0.49\linewidth}
        \centering
        \includegraphics[width=\linewidth]{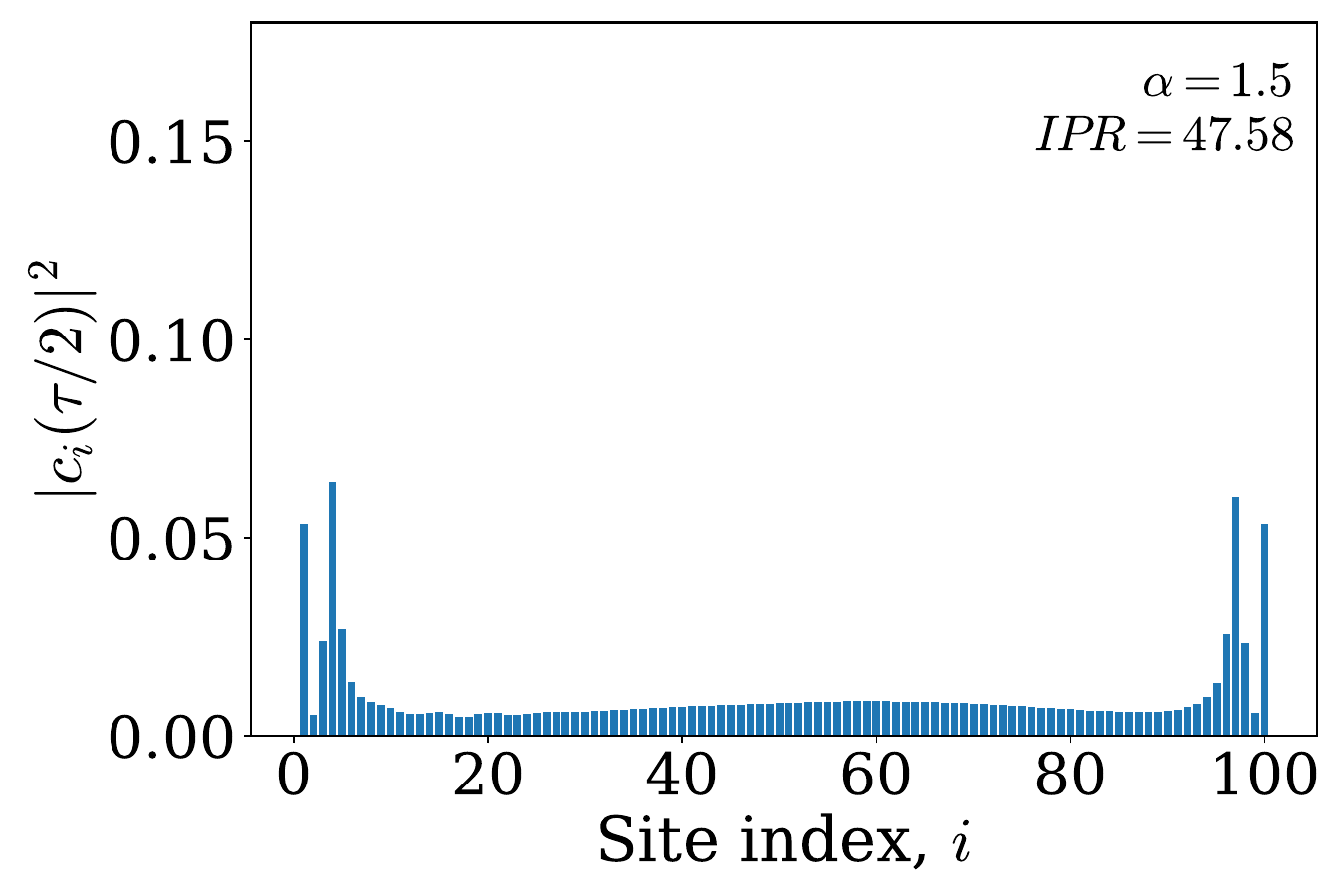}

     \vspace{-0.3cm}

           \includegraphics[width=\linewidth]{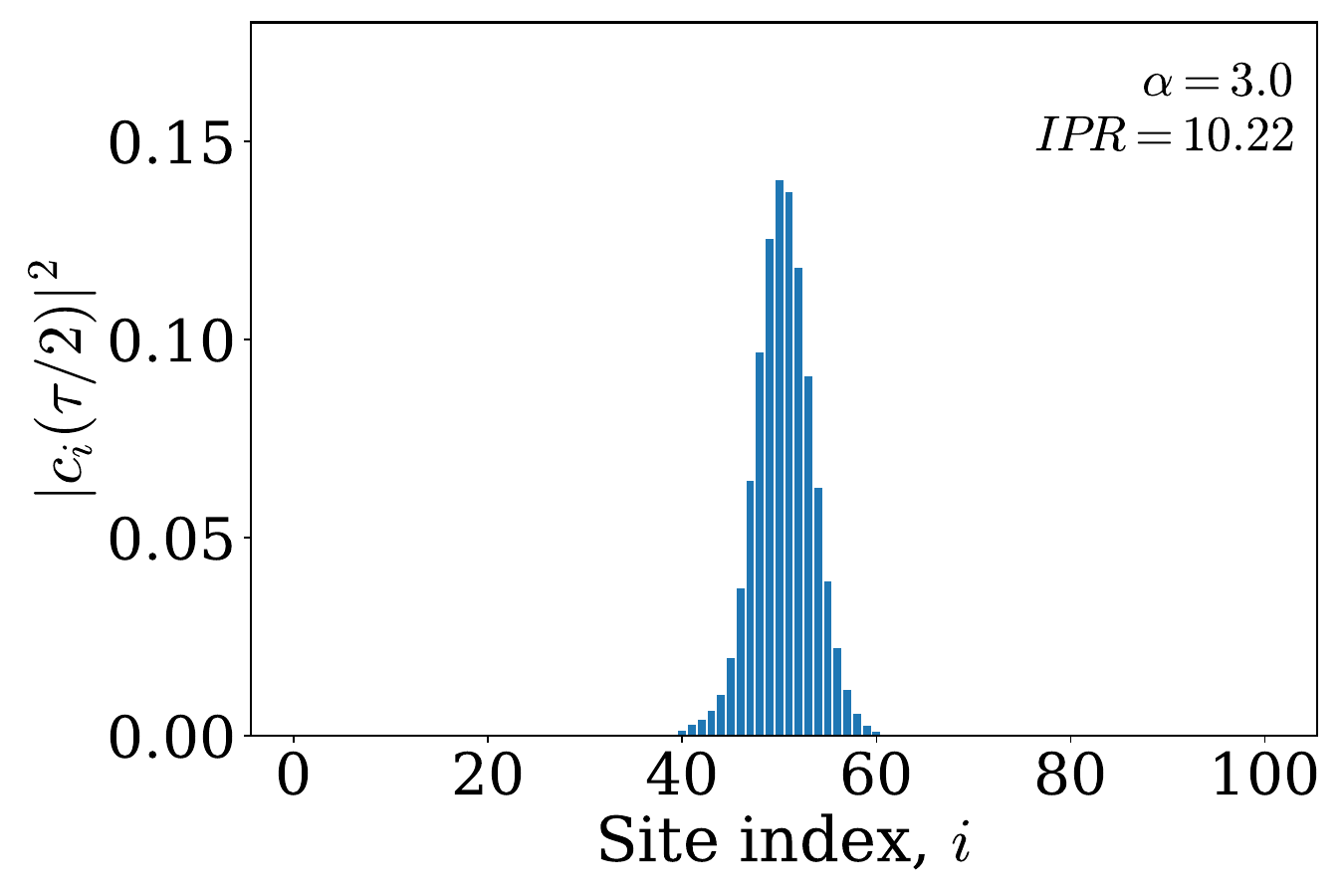}

    \end{minipage}
     \caption{2D snapshots of the transfer fidelity of the excitation to each site for the optimised $N =100-$site chains at $t=\tau_{max}/2$. The IPR (Equation~(\ref{IPR})) value for each snapshot is provided to give a quantitative measure for the wavepacket spread.}
    \label{fig:dynamics-stills}
\end{figure}

Figure \ref{fig:dynamics-stills} shows a snapshot of the transfer fidelity to each site at half the transfer time, $\tau$, for 4 optimised chains. 
There is a marked difference between the snapshots for $\alpha \le 1.5$, showing the probability distribution bunched at the chain sides, and $\alpha\ge 2$, displaying a bell curve at the centre of the chain. 
Writing the state of the system as
$
|\Psi(t)\rangle=\sum_{i=1}^{N}c_i(t)\,|i\rangle,$
where $|i\rangle$ denotes the basis state with the excitation localised at site $i$, the inverse participation ratio (IPR) is defined as \cite{PhysRevA.76.042333,Zwick_2011}
\begin{equation}
\textit{IPR} = \frac{\sum_i |c_i|^2}{\sum_i |c_i|^4}
\label{IPR}\quad,
\end{equation}
which provides a quantitative measure of the delocalisation of the wavepacket. The IPR values shown within Figure~\ref{fig:dynamics-stills} are also evaluated at $\tau/2$, when it should reach its maximal value $IPR\in[1,N].$ The IPR values and distribution snapshots are very similar for $\alpha=3.0$ and $6.0$, hence the latter is omitted. The $\alpha=1.5$ chain has a significantly higher IPR than the rest and a more visually spread-out distribution, indicating a shift in transfer dynamics. This was also accompanied by a change in the adjustments made to the sites, with neighbouring end sites having much larger differences in on-site energies as well as being pulled further away from the bulk for $\alpha<2.0$. For $\alpha=1.0$, the IPR is significantly lower than with $\alpha=1.5$, despite the wavepacket having appreciable components across almost every site in the chain. The probability distribution nevertheless remains more strongly localised at the edges, with lower amplitudes across the intermediate sites. The characteristic bell curve observed for the $\alpha=2.0$ chain is here completely lost. The excitation instead remains largely concentrated on the edge sites, indicating a distinct form of transfer for chains with higher connectivity. 

Our results in Figure~\ref{fig:Fidelity_plot} show that, when normalising the transfer dynamics by the largest coupling in the bulk, the transfer occurs at times much faster than the speed limit in the nearest-neighbour case \cite{Yung_2006}. However, often is the case for the results presented, particularly for $\alpha<3$, that the maximal coupling $J_{max}$ for the optimised solutions actually goes beyond (though never beyond an order of magnitude larger than) the coupling of the bulk, as we allow for the site spacing at the mirror-symmetric edges to come closer than the bulk site spacing. We still normalise with respect to the bulk coupling for the sake of comparison, as this is the only coupling value that is consistent across all $\alpha-$values and chain lengths. We also observe that, for the $\alpha=1$ solutions, the magnetic fields required to induce the few-mode coherent oscillations underlying the exceptionally fast QPST are often well approximated by the largest eigenvalue of the corresponding unperturbed Coulomb-interaction Hamiltonian with zero local field differences. This observation suggests that the field strength required to realize this distinct transfer mechanism may exhibit only weak system-size dependence. This weak system-size dependence is consistent with the modest growth of the largest eigenvalue with $N$ for the $\alpha=1$ Coulomb interaction, implying that substantially larger systems require only a modest increase in the corresponding boundary-field scale. We therefore conjecture that, despite the increasing spectral radius of the unperturbed Coulomb Hamiltonian with system size, the required magnetic fields on the sender and receiver sites grow only modestly and remain experimentally tractable for substantially larger systems.

\label{results}
\subsection{Robustness}
\begin{figure*}[t!]
    \includegraphics[width=0.9\linewidth]{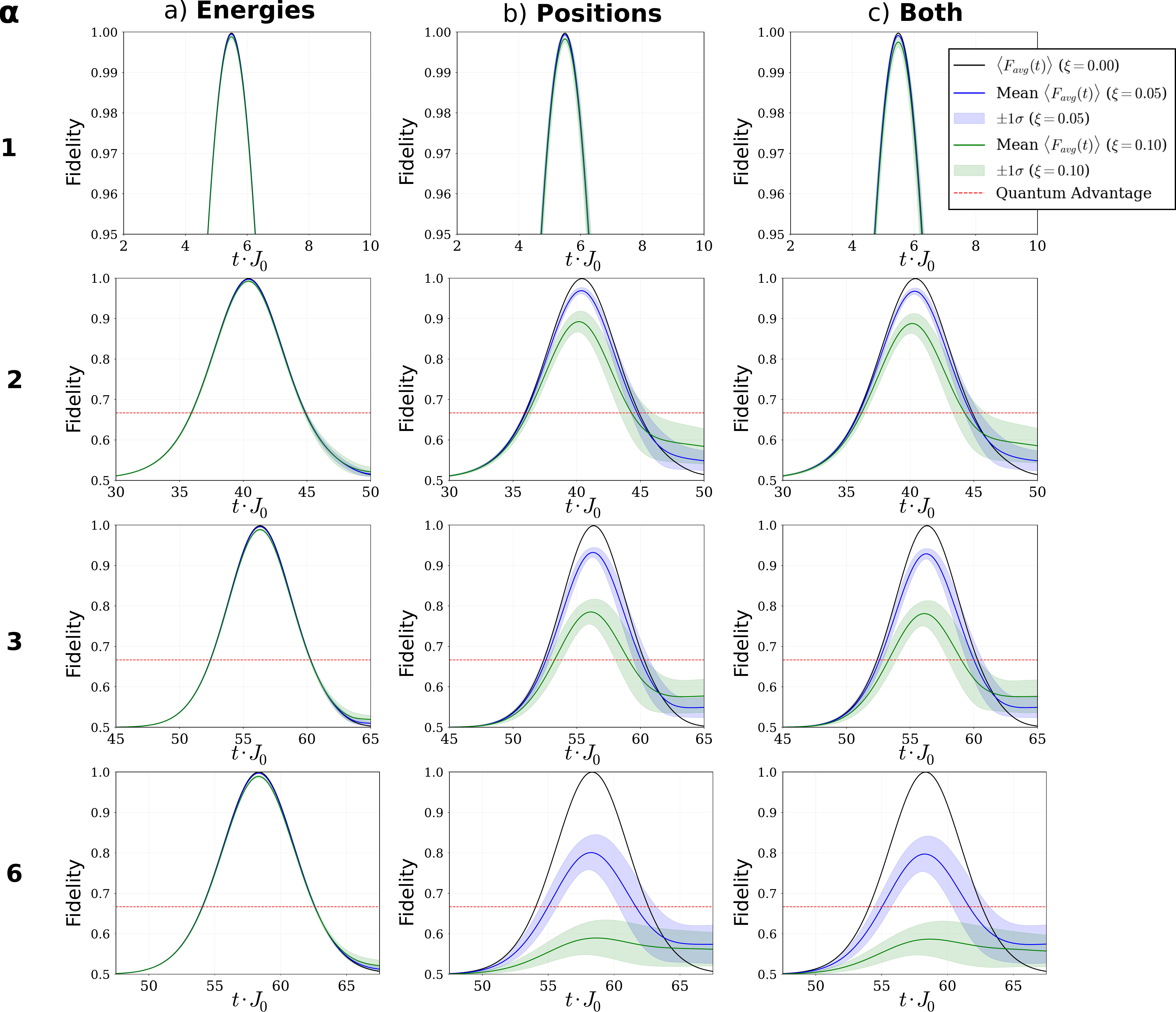}
    \caption{Plots detailing the effect of introducing random disorder into the on-site energies and positions of optimised $N=100$ chains for the 4 interaction strengths of interest. The columns show the effects of perturbing the site energies alone (a), the site positions alone (b), and both simultaneously (c).
 The $\alpha=1$ panels are zoomed-in as the fidelity scale is much more concentrated due to disorder having significantly less effect as other $\alpha$-values.}
    \label{robustness-plot}
\end{figure*}

Furthering the goal of experimental feasibility, we have investigated the effect of random asymmetric disorder (i.e., disorder that does not preserve mirror symmetry) on both the position profile and on-site energies within the optimised LR chain solutions. The on-site energies are modified as
\begin{equation}
\varepsilon_i^{\mathrm{eff}} = \varepsilon_i + \xi \,a_i, \quad i = 1, \ldots, N,
\end{equation}
\begin{equation*}
a_i \sim \mathcal{U}\left[-0.5, 0.5\right],
\end{equation*}
where $\xi$ denotes the disorder strength (e.g., $\xi = 0.1$ corresponds to a 10\% fluctuation amplitude). Similarly, the inter-site distances entering are perturbed according to
\begin{equation}
r_i^{\mathrm{eff}} = r_i \left(1 + \xi b_i \right), \quad i = 1, \ldots, N-1,
\end{equation}
\begin{equation*}
b_i \sim \mathcal{U}\left[-0.5, 0.5\right],
\end{equation*}
where $a_i$ and $b_i$ are independent random variables and $\mathcal{U}$ is the uniform sampling. For the results presented within Figure \ref{robustness-plot}, the effective local magnetic field/on-site energies (Figure \ref{robustness-plot}(a)), and positions (Figure \ref{robustness-plot}(b)) are averaged over 1000 samples, to collect the mean fidelity values and standard error. Figure \ref{robustness-plot}(c) combines the two. The noise simulations presented a significant robustness in the mean transfer fidelity $\geq 96\%$ up to a disorder error strength of $5\%$ ($\xi = 0.05$) for positional noise, and up to $10\%$ ($\xi = 0.10$) for on-site energy noise. It is important to note that asymmetric errors in the range of 5--10\% are relatively large in the context of state-of-the-art trapped-ion and optical-lattice platforms. As the site positioning on quantum hardware can be achieved with high spatial precision \cite{Yu_2019,Wang2020}, the error model considered in Figure~\ref{robustness-plot} should be interpreted as a deliberately conservative scenario, intended to investigate the robustness of the worst-case against imperfections in the positions of the sites. Therefore, the results presented in Figure~\ref{robustness-plot} represent a strict benchmark for the stability of the protocol under experimentally pessimistic conditions. 
\label{R}

Figure \ref{robustness-plot} shows that across all considered interaction strengths, random disorder in the on-site energies has very little effect on fidelity. As displayed in Figure \ref{fig_3}, the on-site energy of the first site is responsible for positioning the initial wavepacket in k-space, whereas the energies of subsequent sites subtly shift the spectrum of the chain. It therefore follows that random disorder of the order $0.1 J_0$ would only have a significant effect on the first site. In the cases of $\alpha=1.0$ and $2.0$, the first on-site energy was $>1 J_0$ for the optimised configurations tested, meaning the disorder strength is smaller in comparison. Also, as shown in Figure \ref{fig_4}, the ordered spectrum for $\alpha=3.0$ has a significantly larger linear section, suggesting that small changes to the magnetic field strengths would not move the wavepacket out of the linear section. This is similarly the case for $\alpha=6.0$.

The effect of random disorder in the positions, however, tells a much more interesting story. As $\alpha$ increases, so too does the effect of disorder to the point that a $10\%$ disorder for $\alpha=6.0$ loses quantum advantage. However, a $5\%$ disorder strength for $\alpha=3.0$ retains $>0.9$ average fidelity to $\pm 1\sigma$, implying a reasonable level of robustness to relatively large positional disorder. For $\alpha=1.0$, there was almost no change to average fidelity, even at $10\%$ disorder strength. Even when simultaneously applying disorder to both positions and on-site energies, the $\alpha=1.0$ configuration retained $>0.99$ average fidelity to $\pm1\sigma$ at $10\%$ disorder strength. Furthermore, for sufficiently large systems ($N \sim 75$–$200$), we found that specific boundary-engineered solutions exhibit a stability under addition of a few sites to the bulk while maintaining nearly identical average fidelity values. This highlights the robustness and adaptability of these solutions, which may be exploited when exact system sizes cannot be realised.
 While the robustness to variations in the on-site energies is consistent with that observed for the other interaction strengths, the pronounced robustness to positional disorder at $\alpha=1.0$ is particularly striking. Not only does the $\alpha=1.0$ regime enable significantly faster transfer, but it also appears to exhibit an intrinsically greater tolerance to positional imperfections. It is possible that this robustness to changes in site positions is due to the higher interaction strength; each site has a stronger connection to surrounding sites and can better average out perturbations. Also, it is possible that this is due to the limited number of active eigenmodes displayed in Figure \ref{fig_4}. Since the majority of the modes are dynamically inactive, perturbations applied to the vast majority of bulk sites would have little effect, resulting in the observed robustness. However, further robustness analysis is beyond the scope of this study.

\section{Conclusions and Outlook}
\label{CO}
We have presented distinct transfer mechanisms within optimised power-law, long-range-interacting spin$-\frac{1}{2}$ systems. While very-high fidelity solutions are found for all power-laws, particularly fast transfer is associated with power exponents $1\le\alpha\le 1.5$, also showing favourable sublinear scaling with chain length. These results are found using a minimalist engineering approach yielding a realistic pathway to experimental implementation and subsequent benchmarking. Furthermore, robustness analyses present, for systems with higher connectivity than an effective nearest-neighbour system, a marked improvement in robustness to static disorder/parameter imprecision. This result further underscores the implementability and potential witnessing of the distinct modes of transfer in real hardware. 
\section*{Acknowledgements}
TJGA and IDA acknowledge funding from the Royal Society under the grant IES\textbackslash R3\textbackslash 243264 - International Exchanges 2024 Global Round 3. C.C. Nelmes acknowledges support from EPSRC, grant number is EP\textbackslash W524657\textbackslash1.
\appendix

\section[\appendixname~\thesection]{}

\begin{figure}[H]
    \centering

    \caption*{(a) 0 optimised sites}\includegraphics[width=0.5\linewidth]{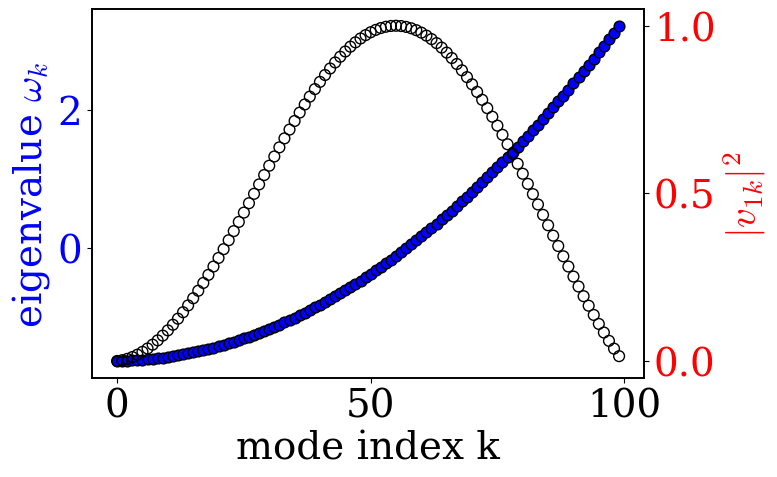}

    \caption*{(b) 1 optimised site}\includegraphics[width=0.5\linewidth]{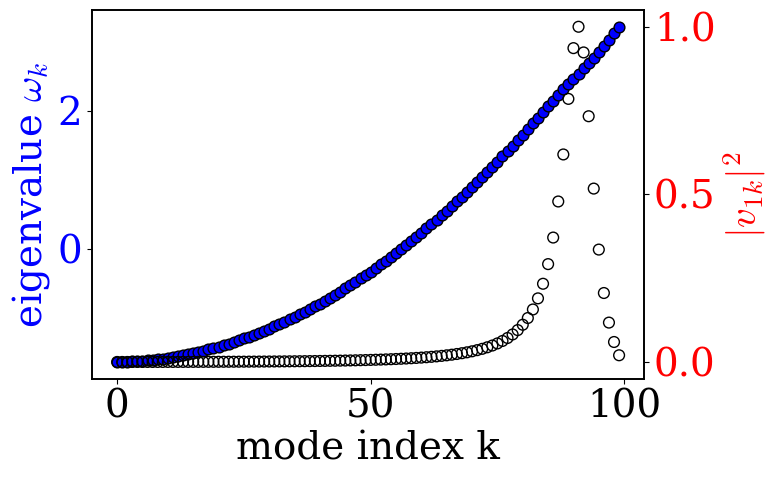}

     \caption*{(c) 2 optimised sites}\includegraphics[width=0.5\linewidth]{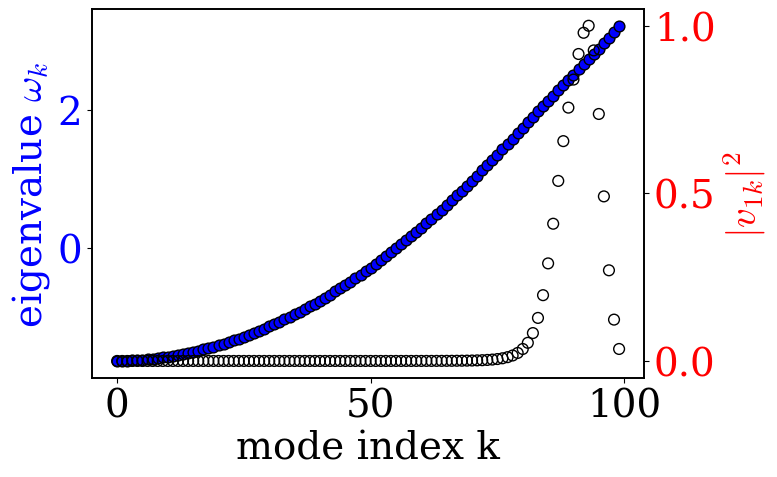}

    \caption*{(d) 3 optimised sites}\includegraphics[width=0.5\linewidth]{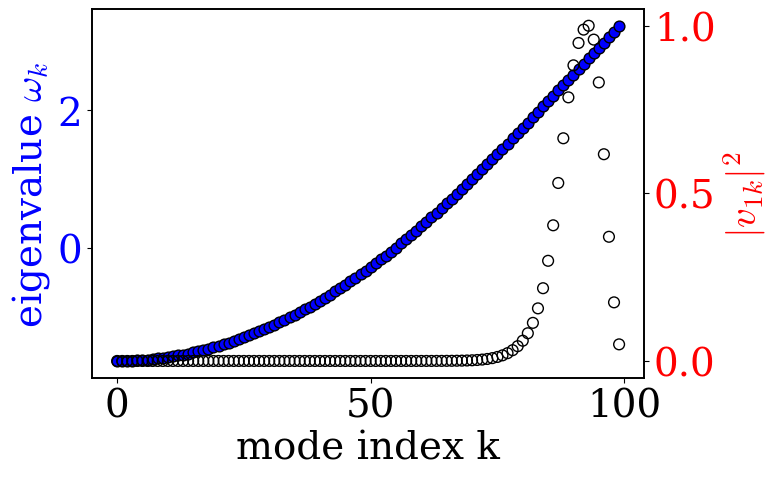}

    \caption{The increasing-ordered spectrum $\omega_k$ and initial wave packet in $k$-space $v_{1k}^2$ across different optimisation configurations. The differences in appearance of the initial wavepacket in $k$-space between (d) and 4 optimised sites is indistinguishable, therefore 4 is omitted.}
    \label{fig_3}
\end{figure}

\begin{table}[H]
\small
\caption{
Maximum average fidelity and corresponding arrival time
$(\max\langle F(t)\rangle,\tau_{\max})$
for several chain lengths $N$ and interpolation parameters $\alpha$.
}
\centering
\renewcommand{\arraystretch}{1.0}
\begin{tabular}{|c|c|c|c|c|c|}
\hline
\multicolumn{6}{|c|}{$\langle F(\tau)\rangle\,;\,\tau$}
\\ \hline

\diagbox{$N$}{$\alpha$}
& $1.0$
& $1.5$
& $2.0$
& $3.0$
& $6.0$
\\ \hline

$25$
& $0.999; 3.15$
& $0.999; 8.99$
& $0.999; 10.1$
& $0.999;15.1$
& $0.999; 16.0$

\\ \hline

$50$
& $0.999; 3.82$
& $0.999;16.9$
& $0.999; 19.7$
& $0.999 ;26.6$
& $0.999; 27.8$

\\ \hline
$75$
& $0.999; 4.50$
& $0.999; 21.8$
& $0.999; 31.4 $
& $ 0.998;40.4$
& $0.999; 43.7$

\\ \hline
$100$
& $0.999; 5.50$
& $0.999 ;25.0$
& $0.999; 40.4$
& $0.998; 56.3$
& $0.999; 58.3$

\\ \hline
$151$
& $0.999; 5.78$
& $0.999; 34.1$
& $0.999; 60.2$
& $ 0.999; 79.76$
& $0.999 ; 80.96$ 

\\ \hline

$200$
& $0.999; 6.01$
& $0.997; 39.5$
& $0.998;75.48$
& $0.998;107.7$
& $0.999; 111.1$

\\ \hline
\end{tabular}
\label{tab:alpha_scan}
\end{table}
Table~\ref{tab:alpha_scan} summarises the maximum average fidelity and corresponding arrival time for the different chain lengths and power-law exponents considered. The maximum average fidelity remains close to unity across all configurations, demonstrating that high-fidelity state transfer can be maintained over a broad range of $N$ and $\alpha$. In contrast, the corresponding arrival time $\tau$ increases systematically with both chain length and $\alpha$. The increase is particularly pronounced for larger $\alpha$, reflecting the progressively shorter-ranged nature of the interactions and the resulting slower propagation of quantum information through the chain. Figure~\ref{fig_3} shows the corresponding eigenvalue spectrum and initial wave-packet weights in $k$-space as the number of optimised boundary sites is increased. While optimisation substantially modifies the spectral structure and wave-packet distribution initially, the differences become progressively less pronounced as more boundary sites are optimised. In particular, the four-site configuration is visually indistinguishable from the three-site case in $k$-space and is therefore omitted for clarity.
\begin{table}[H]
\caption{
Maximum average fidelity and corresponding arrival time
$(\max\langle F(t)\rangle,\tau_{\max})$
for several chain lengths $N$ and power-law exponents $\alpha$.
}
\centering
\renewcommand{\arraystretch}{1.0}
\begin{tabular}{|c|c|c|c|c|c|}
\hline
\multicolumn{6}{|c|}{$\langle F(\tau)\rangle\,;\,\tau$}
\\ \hline

\diagbox{$N$}{$\alpha$}
& $1.0$
& $1.5$
& $2.0$
& $3.0$
& $6.0$
\\ \hline

$25$
& $0.999; 3.15$
& $0.999; 8.99$
& $0.999; 10.1$
& $0.999;15.1$
& $0.999; 16.0$

\\ \hline

$50$
& $0.999; 3.82$
& $0.999;16.9$
& $0.999; 19.7$
& $0.999 ;26.6$
& $0.999; 27.8$

\\ \hline
$75$
& $0.999; 4.50$
& $0.999; 21.8$
& $0.999; 31.4 $
& $ 0.998;40.4$
& $0.999; 43.7$

\\ \hline
$100$
& $0.999; 5.50$
& $0.999 ;25.0$
& $0.999; 40.4$
& $0.998; 56.3$
& $0.999; 58.3$

\\ \hline
$151$
& $0.999; 5.78$
& $0.999; 34.1$
& $0.999; 60.2$
& $ 0.999; 79.76$
& $0.999 ; 80.96$ 

\\ \hline

$200$
& $0.999; 6.01$
& $0.997; 39.5$
& $0.998;75.48$
& $0.998;107.7$
& $0.999; 111.1$

\\ \hline
\end{tabular}
\label{tab:alpha_scan}
\end{table}

\section[\appendixname~\thesection]{}
The optimised boundary parameters corresponding to the solutions presented in Figure~\ref{fig_4} are listed in Table~\ref{solutions}. The results show a clear dependence of the optimal boundary configuration on the interaction exponent $\alpha$. For $\alpha=1.0$ and $1.5$, the optimised parameters exhibit larger deviations from the homogeneous bulk, whereas for increasing $\alpha$ the optimised parameters progressively approach those of the unmodified chain. In particular, several of the boundary fields become negligible or vanish entirely for $\alpha\geq1.5$, while the boundary spacings likewise become closer to unity. This indicates that increasingly short-ranged interactions require less pronounced boundary engineering to achieve the optimised transfer observed in Figures~\ref{fig_4} and~\ref{fig:Fidelity_plot}.
\begin{table}[H]
\caption{optimised boundary parameters for mostly homogeneous $N=100$ long-range chains with four modified boundary sites. The bulk couplings are uniform, with only the boundary spacings $d_i=r_{i,i+1}$ and local fields $B_i$ optimised under various power-law exponent strengths.}
\centering
\begin{tabular}{c c c c c}
\toprule
Parameter & $\alpha=1.0$ & $\alpha=1.5$ & $\alpha=2.0$ & $\alpha=3.0$ \\
\midrule

$r_{1,2}$ & 3.970 & 4.780 & 1.926 & 1.429 \\
$r_{2,3}$ & 0.660 & 1.120 & 1.029 & 1.227 \\
$r_{3,4}$ & 1.736 & 0.696 & 1.119 & 1.105 \\
$r_{4,5}$ & 0.500 & 0.741 & 0.900 & 1.047 \\

\midrule

$B_1$ & 8.644 & 4.450 & 2.616 & 0.769 \\
$B_2$ & 5.770 & 0.000 & 1.713 & 0.537 \\
$B_3$ & 5.051 & 0.000 & 0.000 & 0.235 \\
$B_4$ & 5.503 & 1.690 & 0.000 & 0.113 \\

\bottomrule
\end{tabular}
\label{solutions}
\end{table}
\[\]
\section[\appendixname~\thesection]{}
\label{appendixc}
\begin{figure}[H]
    \centering

    \begin{minipage}[t]{0.49\linewidth}
        \centering
        \includegraphics[width=\linewidth]{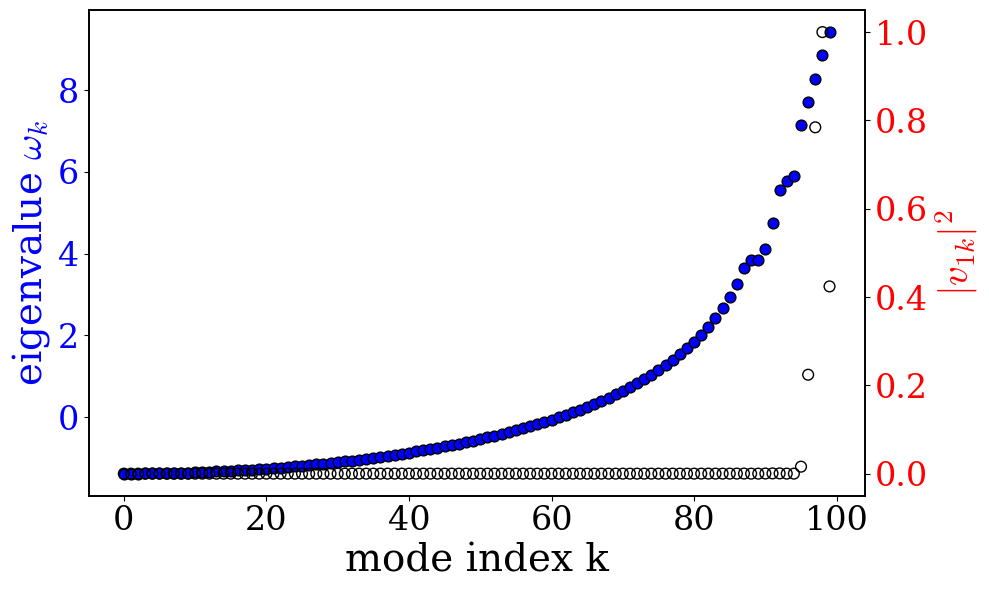}
        \hspace{-1.cm}

        \includegraphics[width=0.9\linewidth]{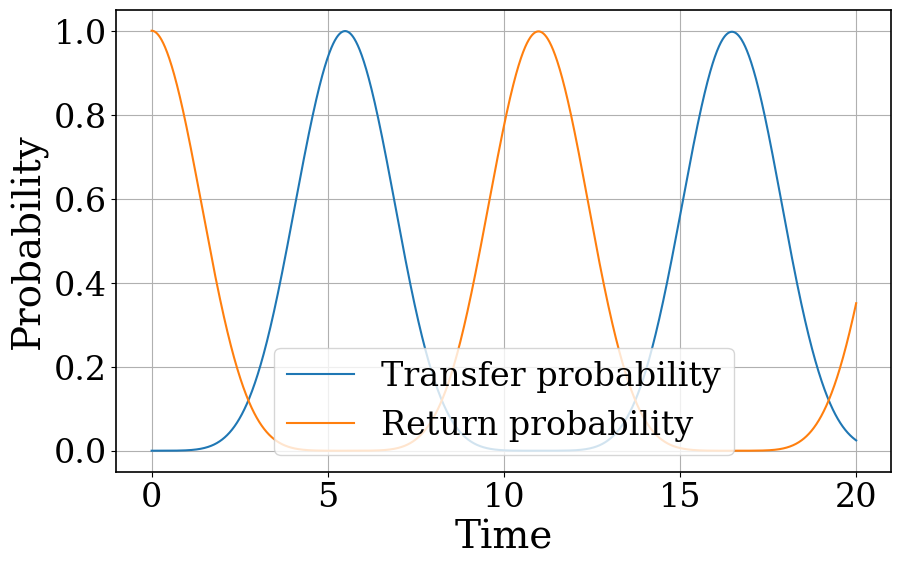}
    
        \hspace{-0.5cm}
        \includegraphics[width=0.95\linewidth]{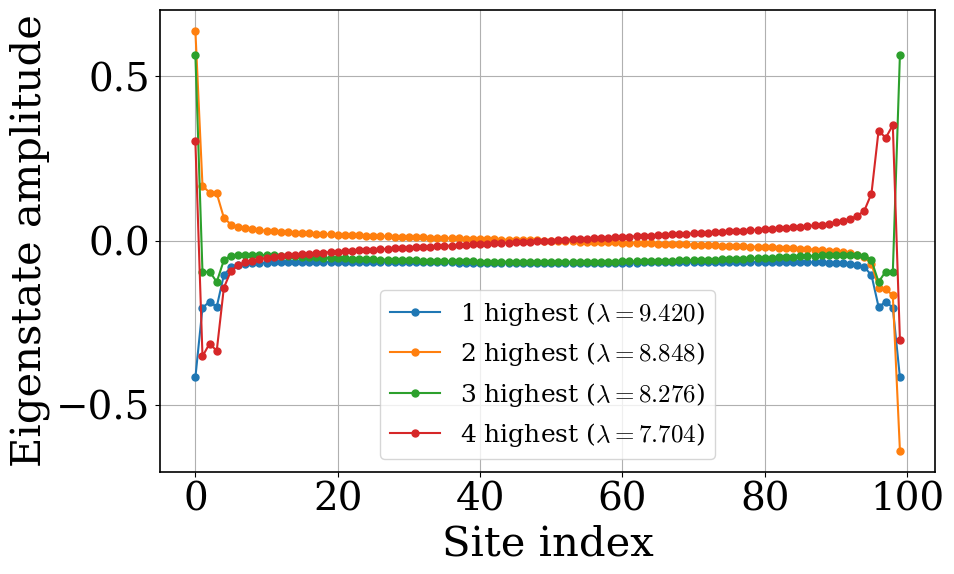}

    \end{minipage}
    \hfill
    \begin{minipage}[t]{0.49\linewidth}
        \centering
        \includegraphics[width=\linewidth]{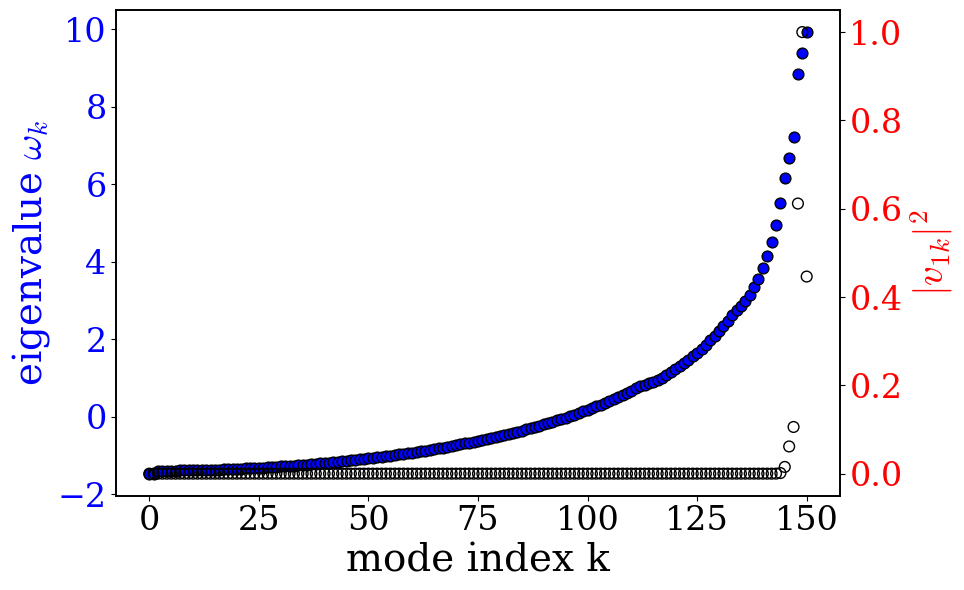}
        \hspace{-1cm}

           \includegraphics[width=0.9\linewidth]{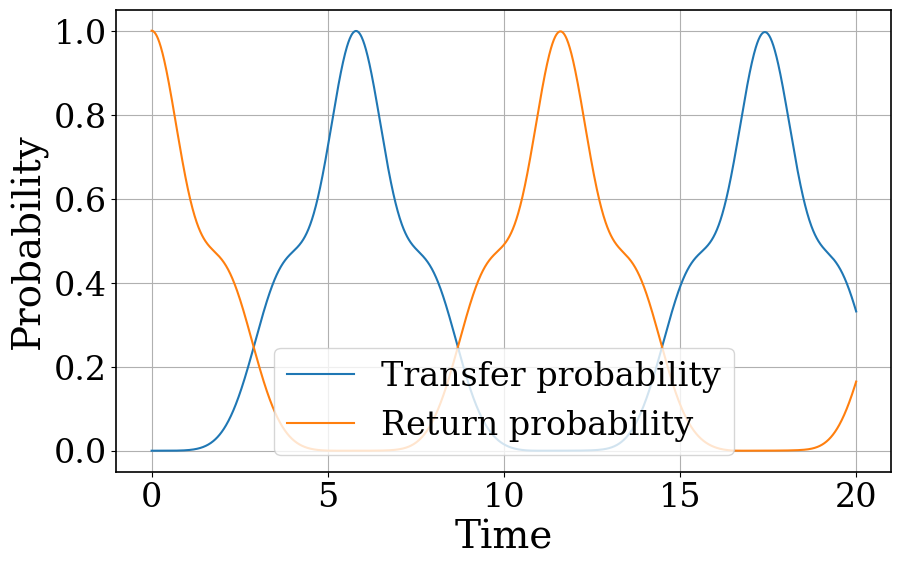}

        \hspace{-0.5cm}
        \includegraphics[width=0.95\linewidth]{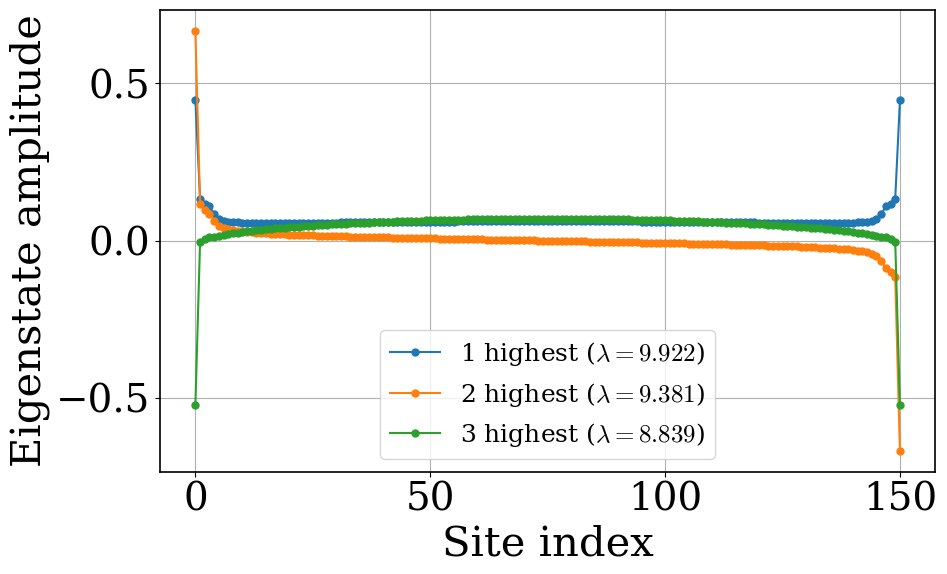}
    \end{minipage}

    \caption{Initial-state localization, transfer-return dynamics, $|f_1^N(t)|^2$ and $|f_1^1(t)|^2$ respectively (Equation~\ref{wavetran}), and the site-basis amplitudes of the high-energy participatory eigenstates for $\alpha = 1$, shown for $N = 100$ (left column) and $N = 151$ (right column).}
    \label{fig_appen2}
\end{figure}

For $\alpha=1.0$, the dynamics exhibit few-mode  coherent oscillations between the end sites. Figure \ref{fig_4} and Figure \ref{fig_appen2} demonstrate that, for $N=100$ and $N=151$  the dynamics involve the highest four and three energy eigenstates, respectively. These dynamics are a little more involved than the simplest form of coherent oscillations involving just two energy eigenstates. Also, the latter simple scenario usually involves two near-degenerate eigenvalues, which, since the transfer time relates to the reciprocal of the energy level separation, inevitably provides very long transfer times. 

The advantage of involving just three or four eigenstates, with equally spaced eigenvalues, is that the transfer time, which again relates to the reciprocal of the separation, is much shorter. As can be seen from the site basis decompositions of the participating eigenstates also shown in the bottom panels of Figure \ref{fig_appen2}, an analogy can be drawn with three- and four-site NN PST chains. With both three and four states, the highest energy eigenstate has even symmetry about the chain centre. The symmetry then oscillates between odd and even as the energy of the participating state decreases. In both examples, a specific superposition of the few participating states corresponds to the excitation completely localised at site $i=1$. At the mirror time, when the odd participating eigenstates have acquired a phase factor of $(-1)$ relative to the even eigenstates, this superposition corresponds to the mirror of the initial state, with the excitation thus completely localised at site $i=N$.

As already stated, the mirror or transfer time relates to the reciprocal of the energy level separation $t=\frac{\pi}{\Delta\omega_{i,i+1}}$, so the transfer can be much faster than oscillations due to two almost degenerate states. Clearly, the solutions presented here do not support mirroring of any multi-qubit initial state, because in each example the whole spectrum is not equally spaced and with oscillating symmetry. Only mirroring of an arbitrary single-qubit quantum state injected at site $i=1$ is to be expected, because that is the problem that the optimisation approach was set up to deliver.

\bibliography{bib}
\end{document}